\documentclass{article}

\usepackage{PRIMEarxiv}

\usepackage[utf8]{inputenc} %
\usepackage[T1]{fontenc}    %
\usepackage{hyperref}       %
\usepackage{url}            %
\usepackage{booktabs}       %
\usepackage{amsfonts}       %
\usepackage{nicefrac}       %
\usepackage{microtype}      %
\usepackage{lipsum}
\usepackage{fancyhdr}       %
\usepackage{graphicx}       %
\usepackage{subcaption}
\usepackage{amssymb}
\graphicspath{{figures/}}     %
\usepackage{amsmath}
\usepackage{placeins}
\usepackage{soul,color}
\usepackage[table,xcdraw]{xcolor}

\usepackage{graphicx}
\usepackage{lipsum} %
\title{Towards Quantitative Evaluation of Crystal Structure Prediction Performance 
}

\author{
  Lai Wei  $^1$\\
 Department of Computer Science and Engineering\\
  University of South Carolina\\
  Columbia, SC 29201 \\
   \And
  Qin Li $^1$\\
 School of big data and statistics\\
  Guizhou University of Finance and Economics\\
  Guizhou, China, 550001 \\
   \And
  Sadman Sadeed Omee \\
 Department of Computer Science and Engineering\\
  University of South Carolina\\
  Columbia, SC 29201 \\
  \And
    Jianjun Hu* \\
 Department of Computer Science and Engineering\\
  University of South Carolina\\
  Columbia, SC 29201 \\
  \texttt{jianjunh@cse.sc.edu} \\
}

\begin{document}
\maketitle

\def\thefootnote{*}\footnotetext{\textit{\underline{Note}: $^1$These authors contributed equally to this work}}\def\thefootnote{\arabic{footnote}}

\begin{abstract}

Crystal structure prediction (CSP) is now increasingly used in the discovery of novel materials with applications in diverse industries. However, despite decades of developments, the problem is far from being solved. With the progress of deep learning, search algorithms, and surrogate energy models, there is a great opportunity for breakthroughs in this area. However, the evaluation of CSP algorithms primarily relies on manual structural and formation energy comparisons. The lack of a set of well-defined quantitative performance metrics for CSP algorithms make it difficult to evaluate the status of the field and identify the strengths and weaknesses of different CSP algorithms. Here, we analyze the quality evaluation issue in CSP and propose a set of quantitative structure similarity metrics, which when combined can be used to automatically determine the quality of the predicted crystal structures compared to the ground truths. Our CSP performance metrics can be then utilized to evaluate the large set of existing and emerging CSP algorithms, thereby alleviating the burden of manual inspection on a case-by-case basis. The related open-source code can be accessed freely at \url{https://github.com/usccolumbia/CSPBenchMetrics}.

\end{abstract}

\keywords{benchmark \and materials discovery \and crystal structure prediction \and distance metric \and performance metrics}

\section{Introduction}

Deep learning-based AlphaFold has been revolutionizing the field of molecular biology by predicting tens of thousands of protein structures from sequences \cite{jumper2021highly}, which can accelerate the understanding of protein structures and functions. However, in the field of materials science, a similar crystal structure prediction problem, which aims to determine the stable structure given only a material composition, remains to be solved. If solved, it can dramatically accelerate the discovery of novel function materials as many material properties such as thermal conductivity, band gap, and elastic constants can be conveniently calculated using first-principle codes such as Density Functional Theory (DFT) based VASP. 

Traditionally, the CSP algorithms are mainly based on the DFT calculation of energies combined with global search algorithms. However, the complexity and demanding computing resources of DFT make it challenging to develop new CSP algorithms. Although the formation energy can be efficiently predicted by graph neural networks (GNNs)~\cite{xie2018crystal,chen2019graph,omee2022scalable}, there is a good amount of accuracy-computing trade-off in this approach. However, recent progress in deep neural network-based energy potentials \cite{chen2022universal} has demonstrated that it is feasible to develop usable CSP algorithms only using neural potentials \cite{cheng2022crystal}. It can be expected that an increasing number of CSP algorithms will emerge as what has happened in the protein structure prediction field with the CASP competitions organized yearly since 1994. In that case, large-scale benchmark studies and objective quantitative evaluation of CSP prediction performances will be needed to illuminate the progress and weaknesses of different algorithms, as has been done in CASP history. 

Currently, there are three main categories of crystal structure prediction algorithms including search-based, template-based, and deep learning-based CSP algorithms. The global search-based CSP algorithms such as USPEX and CALYPSO combine search algorithms with DFT energy calculation for structure search. There are also several open sourced algorithms such as CrySPY \cite{yamashita2021cryspy}, XtalOpt\cite{lonie2011xtalopt}, GASP\cite{tipton2013grand}, and AIRSS \cite{pickard2006high,pickard2011ab}. However, the most widely used and well-established leading software for de novo CSP are GA-based USPEX and particle swarm optimization (PSO)-based CALYPSO. Despite their closed source code, their binary programs can be easily obtained and both come with several advanced search techniques such as symmetry handling, crowding niche, and so on. Global search has also been combined with universal neural potentials for crystal structure prediction as done by the GN-OA algorithm in \cite{cheng2022crystal} and AGOX \cite{christiansen2022atomistic}. With the many possible search algorithms \cite{yin2022search}, the family of such algorithms can keep growing. The second category of CSP algorithms is template-based element substitutions combined with relaxation including TCSP \cite{wei2022tcsp} and CSPML \cite{CSPML}, in which they use rules and machine learning models to guide the selection of structure templates. The last category of CSP algorithms is based on deep learning-based algorithms inspired by the Alphafold. \cite{hu2021alphacrystal}.

With these emerging CSP algorithms, it is critical to benchmark and evaluate their performances in predicting structures of varying complexity so that strengths and obstacles can be identified. However, upon a thorough examination of the relevant literature, it is surprising to find that most of the CSP prediction results are manually verified by authors on a case-by-case basis. This verification process typically involves structure inspection, comparison of formation enthalpies, DFT-calculated energy analysis, examination of property distributions, computation of distances between structures, or a combination of these methods. There has been a severe lack of quantitative measures for objective evaluation of CSP prediction performance. In one of the earliest reports of USPEX (Universal Structure Predictor: Evolutionary Xtallography), which used evolutionary algorithms to identify stable and novel crystal structures that may not be easily discovered through traditional experimental or theoretical methods \cite{glass2006uspex}, the authors compared the energy difference of the predicted structures and ground truth and then compared the structural similarity by manually inspecting the predicted structures against the experimentally determined structures. Similar approaches have been used for verifying predicted crystal structures in related CSP works \cite{oganov2006crystal,oganov2010evolutionary} using evolutionary algorithms. In a related study by Hofmann et al. \cite{hofmann2003crystal}, the authors used the largest distance between the unit cell edges and the nearest grid point of the experimental structure to evaluate CSP performance. Another widely used method for generating predicted crystal structures is CALYPSO (Crystal structure AnaLYsis by Particle Swarm Optimization), which was developed by Wang et al \cite{wang2012calypso}. In this work, energy distributions and the distance against distortion for graphite and diamond structures were utilized to verify the predicted structures. Additionally, in the work by Tong et al. \cite{tong2018accelerating} on accelerating CALYPSO using data-driven learning of a potential energy surface, the authors employed the evolution of the root mean square errors (RMSEs) of the predicted energy and force by the Gaussian Approximation Potential (GAP) for the testing set to evaluate the CALYPSO structure search for a specific cluster. A vector-type structure descriptor distance has also been used for comparing the predicted structures against the ground truths \cite{liu2023shotgun}.

The metrics used in validating the predicted structures against ground truths were usually set by the authors with a certain arbitrariness. So far there is not a set of quantitative indicators of the quality of the predicted crystal structures.  Table \ref{tab:SOTA} provides an overview of the evaluation methods used in state-of-the-art CSP works. The abbreviations M-i, M-o, M-e, M-s, and M-d represent manual structural inspection, comparison with experimentally observed structures, comparison of energy or enthalpy values, success rate analysis, and computation of distances between structures, respectively. It is noteworthy that computational methods such as DFT-energy or enthalpy calculations are commonly employed in many studies. However, manual structural similarity inspection methods continue to be widely used even today, which leads the casual reader to wonder how exactly a predicted crystal structure is evaluated in terms of its prediction quality, especially when the predicted structure does not exactly match the ground truth. Additionally, energy or enthalpy calculations for structure similarity evaluation using DFT can be time-consuming. Furthermore, performance evaluation methods such as success rates, and ad hoc distance calculations between structures present challenges in standardizing, validating, and comparing the CSP results.

\begin{table}[]
\centering
\caption{Overview of state-of-the-art CSP works for validating predicted structures. Abbreviations M-i, M-e, M-s, M-d stand for different validation methods, where M-i, M-e, M-s, M-d represent manual inspection, comparison of the energy or enthalpy, success rate, and computation of distances between structures.}
\begin{tabular}{|l|l|l|l|l|l|l|}
\hline
\textbf{Author} & \textbf{Algorithm} & \textbf{Year} & \textbf{M-i} & \textbf{M-e} & \textbf{M-s} & \textbf{M-d} \\ 
\hline
Hofmann DWM \cite{hofmann2003crystal}            & Data Mining                       & 2003                              & \checkmark                                  &                   &                   &                   \\
Scott M. Woodley \cite{woodley2004structure}       & Evolutionary Algorithm            & 2004                              &                                      & \checkmark                 & \checkmark                 &                   \\
R. Oganov \cite{glass2006uspex}            & Evolutionary Algorithm            & 2006                              & \checkmark                                   & \checkmark                 &                   &                   \\
R. Oganov \cite{oganov2006crystal}             & Evolutionary Algorithm            & 2006                              & \checkmark                                   & \checkmark                 &                   &                   \\
Christopher C. Fischer \cite{fischer2006predicting} & Data Mining                       & 2006                              & \checkmark                                    &                   &                   &                   \\
Kuo Bao \cite{bao2009structure}                & Hopping method                    & 2009                              &                                      & \checkmark                 &                   &                   \\
Giancarlo Trimarchi \cite{trimarchi2009predicting}    & Evolutionary Algorithm            & 2009                              & \checkmark                                   & \checkmark                 &                   &                   \\
R. Oganov \cite{oganov2010evolutionary}              & Evolutionary Algorithm            & 2010                              &                                      & \checkmark                 &                   &                   \\
David C. Lonie \cite{lonie2011xtalopt}        & Evolutionary Algorithm            & 2011                              &                                      & \checkmark                 &                   &                   \\
Yanchao Wang \cite{wang2012calypso}          & Particle swarm optimization (PSO) & 2012                              &                                      & \checkmark                 &                   & \checkmark                 \\
S Q Wu \cite{wu2013adaptive}                & Evolutionary Algorithm            & 2013                              & \checkmark                                    &                   &                   &                   \\
Anton O. Oliynyk \cite{oliynyk2017disentangling}      & Data-Driven: ML                   & 2017                              &                                      & \checkmark                 &                   &                   \\
Qunchao Tong \cite{tong2018accelerating}          & Particle swarm optimization (PSO) & 2018                              &                                      & \checkmark                 &                   & \checkmark                 \\
Maximilian Amsler \cite{amsler2010crystal}      & Hopping method                    & 2018                              &                                      & \checkmark                 &                   &                   \\
Asma Nouira \cite{nouira2018crystalgan}           & Data-Driven: ML                   & 2018                              &                                      & \checkmark                 & \checkmark                 &                   \\
Evgeny V. Podryabinkin \cite{podryabinkin2019accelerating} & Evolutionary Algorithm            & 2019                              & \checkmark                                    &                   &                   &                   \\
Lai Wei \cite{wei2022tcsp}               & Template-Based Substitution       & 2022                              &                                      &                   &                   & \checkmark                 \\
Xuecheng Shao \cite{shao2022symmetry}         & A symmetry-orientated method      & 2022                              & \checkmark                                    &                   & \checkmark                 &                   \\
Xiangyang Liu \cite{liu2021copex}         & Evolutionary Algorithm            & 2022                              & \checkmark                                    &                   &                   & \checkmark                 \\
Yanchao Wang \cite{wang2022crystal}          & Particle swarm optimization (PSO) & 2022                              & \checkmark                                 &                   &                   & \\
Guanjian Cheng \cite{cheng2022crystal}          & Data-Driven: ML & 2022                              & \checkmark                                   &                   & \checkmark                  &\\
\hline
\end{tabular}

\label{tab:SOTA}
\end{table}

Inspired by the variety of quantitative metrics used in evaluating molecule generation algorithms by the benchmark MOSES \cite{polykovskiy2020molecular}, here we aim to address the challenges in defining good structure distance/similarity scores to measure the quality of CSP algorithms. We evaluated a series of energy and structure-based performance metrics for crystal structure prediction algorithms. For each metric, we check how their values correlate with the formation energy differences and perturbation deviations between the predicted structures and the ground truth structures. We tested their correlations for both random perturbations (applied to each atomic site independently) and symmetric perturbations \cite{shao2022symmetry} (applied only to Wyckoff sites without disrupting the symmetry), both of which have been adopted in CSP algorithms. We also showed that while every single metric cannot be used to fully characterize the quality of a predicted structure against the ground truth, together they can capture the key structural similarity. Applications of these metrics were additionally used to compare the performance of CSP algorithms based on different search algorithms. We have also used these metrics to visualize the search trajectories of the structures for the GN-OA algorithms and explained its key limitations.

\section{Method}
\label{sec:headings}

\subsection{Evaluation metrics}
Evaluation metrics play a crucial role in materials science research, as they provide a quantitative way to assess the performance and effectiveness of different material structure prediction algorithms. Currently, there are many evaluation benchmark metrics in the molecule research area, such as RDKit \cite{landrum2013rdkit} and MOSES \cite{polykovskiy2020molecular}. However, in the field of materials informatics, we don't have a unified standard for evaluating the similarity between two crystal structures that arise during the crystal structure prediction process. Here we introduce a set of benchmark metrics for CSP by combining the energy distance along with several common distance metrics, including M3GNet energy distance, minimal Wyckoff RMSE distance, minimal Wyckoff MAE distance, RMS distance, RMS anonymous distance, the Sinkhorn distance, the Chamfer distance, the Hausdorff distance, superpose RMSD distance, edit graph distance, XRD spectrum distance, fingerprint distance to standardize the comparison of material structure prediction algorithms. 

The structure similarity in CSP has a unique property: the candidate structure and the ground truth structure compared have the same number of atoms within the given unit cell. Then the key step is to match atoms of one structure to the corresponding atoms of the other structure to minimize the MAE error. 
There are several desirable characteristics for a good structure similarity measure: 1) correlation: the structure difference should correlates well with the distance metric; 2) convergence: when the predicted structures during the CSP search approach to the ground truth, the distance metric scores should converge to 0; 3) applicability: the distance metric should be used to not just evaluate very similar structures, but also relative distant intermediate structures, which lacks in the success rate metric. Here we introduce eleven distance metrics that may be used in evaluating the prediction performance of CSP algorithms.

\subsubsection{Energy Distance (ED)}
The formation energy is the energy required to form a material from its constituent elements in their reference states, which can provide information regarding the stability and reactivity of materials. While DFT calculation of formation energy is ideal for accuracy, it is too slow in many applications. As a result, machine learning-based energy models have become an important topic with significant progress recently for materials discovery and design. Here we use the M3GNet \cite{chen2022universal}, a graph neural network-based surrogate potential model, to calculate the formation energies of the ground truth structure and the predicted structure and then their energy distance. The formula is shown in the following equation:
\begin{equation}
    \mathrm{ED} = |E_p - E_g| 
    \label{eqn:eq_ED}
\end{equation}, where $E_p$ is the energy of the predicted structure while the $E_g$ is the ground truth structure.

\subsubsection{Wyckoff position fraction coordinate distance (WD)}

The Wyckoff position fraction coordinate distance is used to compare two structures that have the same Wyckoff position configurations in the symmetrized structures. It is useful to measure the similarity of the candidate structures and the ground truth structures for those CSP algorithms that can search structures while preserving symmetry (space groups). 

RMSE stands for Root Mean Square Error, which can be calculated as the square root of the average of the squared differences between the predicted and actual values.

\begin{equation}
    \mathrm{WD_{RMSE}} = \sqrt{\frac{1}{n}\sum_{i=1}^{n}(y_i - \hat{y}_i)^2} 
    \label{eqn:eq_RMSE}
\end{equation}
where variables $y$ and $\hat{y}$ representing the actual and predicted values, respectively. To calculate this distance, the input structures have to be symmetrized first e.g. using Pymatgen's SpaceGroupAnalyzer module. 

Minimal MAE Distance is the minimum mean absolute error (MAE) distance between two sets of data points. It is a measure of the closeness between two sets of data points. The MAE distance is calculated by taking the absolute difference between corresponding data points in the two sets of data, summing these absolute differences, and dividing by the total number of data points.
Let $X = \{x_1, x_2, \dots, x_n\}$ and $Y = \{y_1, y_2, \dots, y_n\}$ be two sets of $n$ data points. The mean absolute error (MAE) distance between $X$ and $Y$ is defined as:

\begin{equation}
\mathrm{WD_{MAE}} = \frac{1}{n} \sum_{i=1}^{n} |x_i - y_i|
\end{equation}

The minimal MAE distance is the smallest possible MAE distance that can be obtained between $X$ and $Y$ by permuting the data points in $X$ and $Y$.

\subsubsection{Adjacency Matrix Distance (AMD)}

The adjacency matrix (M) is widely used to represent the connection topology of atoms for a given crystal structure. The value of $M(i,j)$ is set to 1 if there exists a bond between the atom $i$ and atom $j$ or set to 0 if otherwise. Here we use the canonical distance as the cutoff distance for a pair of element types 
to define the connection status. Given two structures $S_1$ and $S_2$ with adjacency matrices $M_1$ and $M_2$, the adjacency matrix distance is defined as:

\begin{equation}
\mathrm{AMD} = 1-\frac{2 n}{n_1+n_2}
\end{equation} where $n$ is the number of matrix cells that both matrices have the value of 1. $n_1$ is the number of matrix cells of $M_1$ with the value of 1 and $n_2$ is the number of matrix cells of $M_2$ with the value of 1.

The AMD can be used to measure the topological similarity between two compared structures, which are assumed to have an equal number of atomic sites. However, we found that the correlation between the perturbation magnitude and the AMD distance is weak (see Supplementary Figure $S3$ and $S4$).

\subsubsection{Pymatgen RMS distance (PRD) and RMS Anonymous Distance (PRAD)}

We calculate RMS Anonymous Distance using the structure\_matcher module of the PyMatGen (Python Materials Genomics) package \cite{ong2013python}, which allows distinct species in one structure to map to another. It also calculates the root-mean-square error (RMSE) between two structures according to equation \ref{eqn:eq_RMSE}, but its atomic site matching process does not consider the difference of atom types before calculating the RMSD. It is useful in cases where one wants to compare the overall structural similarity between two structures, without being concerned with the differences in atom types.

\subsubsection{Sinkhorn Distance (SD)}
The Sinkhorn Distance (SD)~\cite{cuturi2013sinkhorn} is a kind of distance metric that is used to compare two probability distributions or point clouds in high-dimensional spaces. 
To represent two crystal structures as point cloud, we can consider their atomic sites. Given two structures as $S_1 = \{p_1, p_2, \dots, p_m\}$ and $S_2 = \{q_1, q_2, \dots, q_n\}$ where $p_i$ and $q_i$ are atomic sites of structure $S_1$ and $S_2$, respectively, we can formally define SD as:

\begin{equation}\label{eq4}
    SD\left(S_1, S_2\right) = \frac{1}{\epsilon}\Bigg(\sum_{i,j} T_{i,j}\log\frac{T_{i,j}}{u_iv_j} - \log C\Bigg)
\end{equation}

In Eq.~\ref{eq4}, $S_1$ and $S_2$ are the crystal structures to be compared, and $u$ and $v$ denote the marginals of $S_1$ and $S_2$ which represent the total mass at each point (atomic site) in each distribution, respectively, $\epsilon$ denotes a regularization parameter, $C$ is a normalization constant and $T$ is the transport plan which can be computed via the following equation:

\begin{equation}\label{eqt}
    T_{i,j} = \exp(-\epsilon h_{i,j})u_iv_j
\end{equation}

In Eq.~\ref{eqt}, $h_{i,j}$ denotes the cost of transporting one unit of mass from site $i$ in $S_1$ to site $j$ in $S_2$.

SD can be thought of as a regularized version of the Earth Mover's Distance (EMD), also referred to as the Wasserstein Distance, where the smoothness of the transport plan is controlled by the regularization parameter $\epsilon$. The transport plan gets quite sparse and the SD approaches the EMD when $\epsilon$ is very large. The transport plan becomes very dense and the SD approaches the Euclidean Distance when $\epsilon$ is very tiny.

\subsubsection{Chamfer Distance (CD)}
The Chamfer Distance (CD)~\cite{fan2017point} is defined as the average distance of the summed-up squared distances between two point clouds' nearest neighbor correspondences. Similar to SD, we also represent two crystal structures as atomic sites to define it. Given two structures as $S_1 = \{p_1, p_2, \dots, p_m\}$ and $S_2 = \{q_1, q_2, \dots, q_n\}$ where $p_i$ and $q_i$ are atomic sites of structure $S_1$ and $S_2$, respectively, we can formally define CD as:

\begin{equation}\label{eq5}
    CD\left(S_1, S_2\right) = \frac{1}{m} \sum_{p \in S_1} \min _{q \in S_2}\|p - q\|_2+\frac{1}{n} \sum_{q \in S_2} \min _{p \in S_1}\|q - p\|_2
\end{equation}
In Eq.~\ref{eq5}, $\|p - q\|^2$ is the squared Euclidean Distance between sites $p$ and $q$.
It is relatively fast and easy to compute, can handle large point sets, and is less sensitive to outliers and noise in the data. However, it also has some drawbacks, such as being dependent on the metric used to measure distances between points and being insensitive to the relative ordering of the points.

\subsubsection{Hausdorff Distance (HD)}

The Hausdorff Distance (HD)~\cite{huttenlocher1993comparing} measures the maximum distance between any point in one set and its nearest point in the other set, or vice versa. We represent two crystal structures as atomic sites to define them. Given two structures as $S_1 = \{p_1, p_2, \dots, p_m\}$ and $S_2 = \{q_1, q_2, \dots, q_n\}$ where $p_i$ and $q_i$ are atomic sites of structure $S_1$ and $S_2$, respectively, we can formally define HD as:

\begin{equation}\label{eq6}
    HD(S_1, S_2) = \operatorname*{max} \Big\{\operatorname*{sup}_{p \in S_1} \operatorname*{inf}_{q \in S_2} \|p - q\|, \operatorname*{sup}_{q \in S_2} \operatorname*{inf}_{p \in S_1} \|q - p\|\Big\}
\end{equation}
In Eq.~\ref{eq6}, $\|p - q\|$ is the distance between sites $p$ and $q$, which can be any distance metric such as the Euclidean Distance, the Manhattan Distance, or the Minkowski Distance. The $\sup$ function takes the supremum or the least upper bound of the distances overall points in the set, and the $\inf$ function takes the infimum or the greatest lower bound of the distances overall points in the other set. The $\max$ function takes the maximum of the two supremum, which ensures that the Hausdorff distance is a symmetric metric.

\subsubsection{Superpose Distance (SPD)}

The Superpose Distance (SPD) is a measure of the structural similarity between two 3D protein structures. SPD is essentially a variation of the RMSE, which is a commonly used metric for quantifying the structural similarity between two protein structures.
We again represent two crystal structures as atomic sites to define them.  Given two structures as $S_1 = \{p_1, p_2, \dots, p_n\}$ and $S_2 = \{q_1, q_2, \dots, q_n\}$ where $p_i$ and $q_i$ are atomic sites of structure $S_1$ and $S_2$, respectively.
We use the Superpose3D~\cite{2022GitHub} package which takes these two structures as input representing two sets of atomic sites of the same length $N$. It attempts to superimpose them using rotations, translations, and optionally scale transformations while treating them as rigid objects in order to reduce the RMSE across corresponding sites. The RMSE of the paired sites is calculated as SPD following alignment. It can be defined by the following equation:

\begin{equation}\label{eqspd}
SPD=\sqrt{\frac{1}{N} \sum_{n=1}^N \sum_{i=1}^{s_l}\left|S_{1_{n i}}-\left(\sum_{j=1}^{s_l} h R_{i j} S_{2_{n j}}+T_i\right)\right|^2}    
\end{equation}

In Eq.~\ref{eqspd}, $s_l$ denotes the dimension of the atomic sites, $R_{i j}$ denotes rotation matrix (a $s_l \times s_l$ sized array representing the rotation), where $|R|=1$,  $T_{j}$ denotes a translation vector, and $h$ denotes a scalar.
One limitation of this distance measure is that their superimposing/alignment does not consider the atomic types of the points. 

\subsubsection{Graph Edit Distance (GED)}
The Graph Edit Distance (GED)~\cite{sanfeliu1983distance} is a distance metric used to compare two graphs with possibly different numbers of nodes and edges. It measures the minimum number of operations needed to convert one graph into another. The permitted operations are insertions, deletions, and substitutions of nodes and edges.
GED is defined by the following equation:
\begin{equation}\label{eqg}
    GED (G_1, G_2) = \min_\zeta \Big(\sum_{i\in V_1}\alpha_\zeta(i) + \sum_{(i,j)\in E_1} \beta_\zeta(i,j)\Big)
\end{equation}

In Eq.~\ref{eqg}, $G_1 = (V_1, E_1)$ and $G_2 = (V_2, E_2)$ are the two input graphs to be compared, $\zeta$ is a graph edit path that maps nodes and edges from $G_1$ to $G_2$, $\alpha$, and $\beta$ are the cost of editing a node and an edge, respectively.

Here, we use a measure of the similarity between two crystal structures represented as graphs based on the differences in the connectivity and bonding patterns of the atoms in the structures. We use the atomic number difference as the node substitution cost and the length difference of two edges as the edge substitution cost. Then the GED uses the linear sum assignment algorithm, also known as the Hungarian algorithm \cite{jonker1986improving} to find an assignment of nodes in structure A to nodes of structure B that minimizes the total substitution cost. The algorithm works by iteratively constructing a dual feasible solution, which is then used to generate an alternating path in a bipartite graph. This alternating path is used to update the current assignment and improve the overall cost. We use the implementation in \cite{GED} package for this distance calculation.

This GED distance considers the atomic site element types during its site alignment process, which can complement the shortcoming of the Superpose distance SPD. However, we found that the correlation between the perturbation magnitude and the GED distance is weak (see Supplementary Figure S1 and S2).

\FloatBarrier

\subsubsection{X-ray Diffraction Spectrum Distance (XD)}
X-ray diffraction (XRD) is a technique used to determine the atomic and molecular structure of a material by analyzing the diffraction patterns resulting from X-ray interactions with a crystal. To quantify the similarity between two material structures based on their XRD features, we calculate the Euclidean distance, referred to as the X-ray diffraction (XRD) Spectrum Distance.

Given two crystal structures $S_1 = (p_{x_1}, p_{x_2}, \ldots, p_{x_n})$ and $S_2 = (q_{x_1}, q_{x_2}, \ldots, q_{x_n})$, where $p_{x_i}$ and $q_{x_i}$ represents the XRD features of structures $S_1$ and $S_2$, respectively, in an $n$-dimensional space, the Euclidean XRD spectrum distance $XD$ between the two structures can be represented as follows: 

\begin{equation}
\label{euclidean1}
 XD\left(S_1, S_2\right)   = \sqrt {\sum _{i = 1}^{n}  \left( q_{x_i} - p_{x_i}\right)^2 } 
\end{equation}

\subsubsection{Orbital Field Matrix Distance (OD)}
The Orbital Field Matrix (OFM) is calculated for each site in the supercell by considering the valence shell electrons of neighboring atoms. Each crystal structure is transformed into a supercell, which prevents sites from coordinating with themselves. And then the average of the orbital field matrices at each site is found to characterize a structure. The Orbital Field Matrix (OFM) distance is determined by calculating the Euclidean distance between the OFM features of two structures. 

Given two crystalline structures $S_1 = (p_{o_1}, p_{o_2}, \ldots, p_{o_n})$ and $S_2 = (q_{o_1}, q_{o_2}, \ldots, q_{o_n})$, where $p_{o_i}$ and $q_{o_i}$ represents the OFM features of structures $S_1$ and $S_2$, respectively, in an $n$-dimensional space, the Euclidean OFM distance $OD$ between two structures can be defined as the following equation: 

\begin{equation}
\label{euclidean2}
 OD\left(S_1, S_2\right)   = \sqrt {\sum _{i = 1}^{n}  \left( q_{o_i} - p_{o_i}\right)^2 } 
\end{equation}

\subsection{Evaluation procedure}

We used three ways to evaluate the utility of the selected distance metrics for CSP study including: 1) studying how the structure distance metrics change with the structure perturbation; 2) comparing the CSP performances of three CSP algorithms over a set of test structures; 3) understanding the search dynamics or behavior of the optimization algorithms using the trajectories of the search process.

\section{Results}
\label{sec:others}

\subsection{Evaluation of performance metrics}

To evaluate how different performance metrics reflect the actual closeness of the predicted structures to the ground truth structures, we use two perturbation methods to generate two sets of perturbed crystal structures with varying magnitudes for a given stable structure. We then calculate how their formation energy differences correlate with the performance metric distances as well as the perturbation magnitudes. The first perturbation method directly changes the coordinates of all sites by a random uniform $p$ percentage (from 1 to 50\% with 1000 points) without considering the space group symmetry. %
So the resulting structures may lose thier space symmetry. By perturbing a given stable crystal structure with an increasing series of perturbation magnitudes, we can simulate the search process of CSP algorithms to a certain degree. 
The second perturbation method comes from the PyxTal package \cite{fredericks2021pyxtal}, which can do two types of perturbations over a given structure: one is changing the lattice parameters by a given percentage; the other is perturbing the atomic coordinates of the Wyckoff sites with a given magnitude in \AA. Here we focus on this symmetry-preserving Wyckoff site coordinate perturbation with 2\% lattice perturbation.

\begin{figure}[ht!] 
 \begin{subfigure}[t]{0.32\textwidth}
        \includegraphics[width=\textwidth]{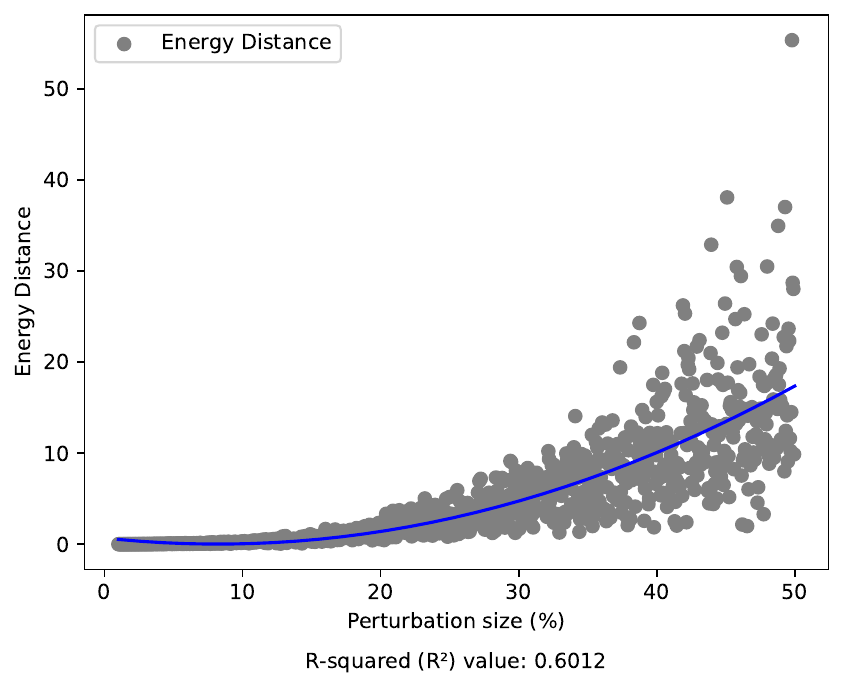}
        \caption{SrTiO\textsubscript{3} (Energy Distance)}
        \vspace{-3pt}
        \label{fig:SrTiO3_target}
    \end{subfigure}
    \begin{subfigure}[t]{0.32\textwidth}
        \includegraphics[width=\textwidth]{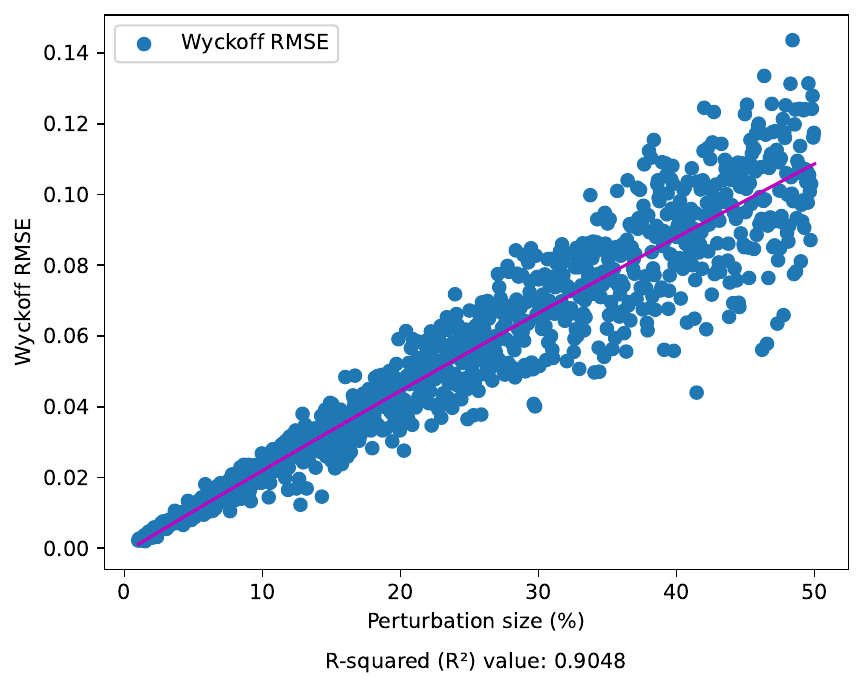}
        \caption{Wyckoff RMSE}
        \vspace{-3pt}
        \label{fig:Ni3S4_target}
    \end{subfigure}    
    \begin{subfigure}[t]{0.32\textwidth}
        \includegraphics[width=\textwidth]{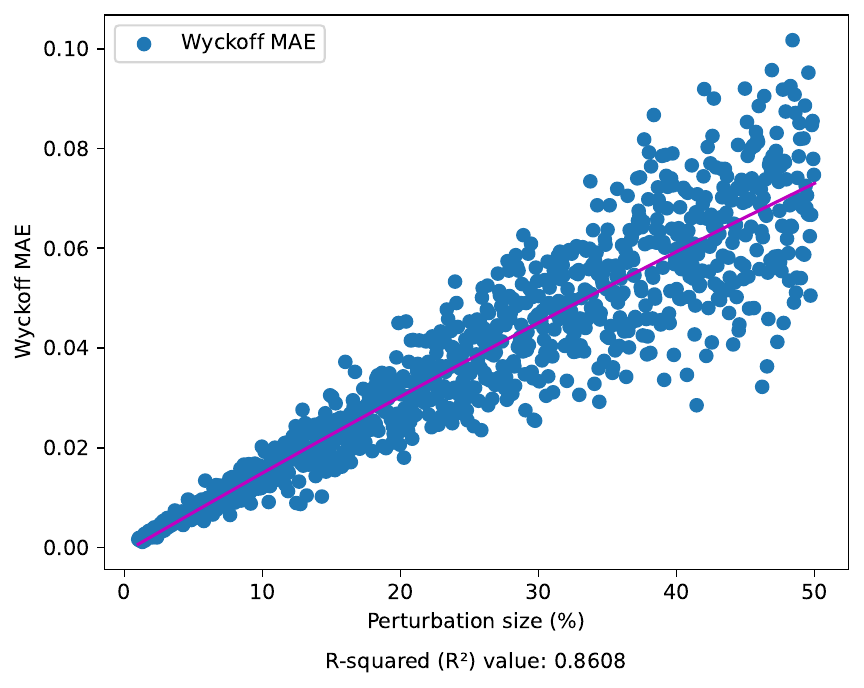}
        \caption{Wyckoff MAE}
        \vspace{-3pt}
        \label{fig:NiS2_target}
    \end{subfigure}    
    \begin{subfigure}[t]{0.32\textwidth}
        \includegraphics[width=\textwidth]{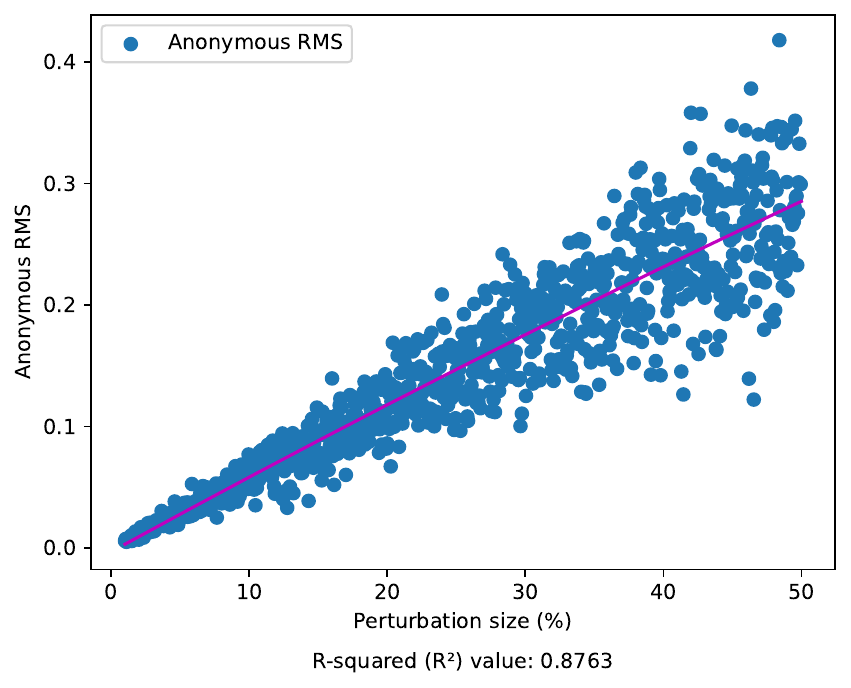}
        \caption{Anonymous RMS}
        \vspace{-3pt}
        \label{fig:Ni3S4_predict}
    \end{subfigure}
        \begin{subfigure}[t]{0.32\textwidth}
        \includegraphics[width=\textwidth]{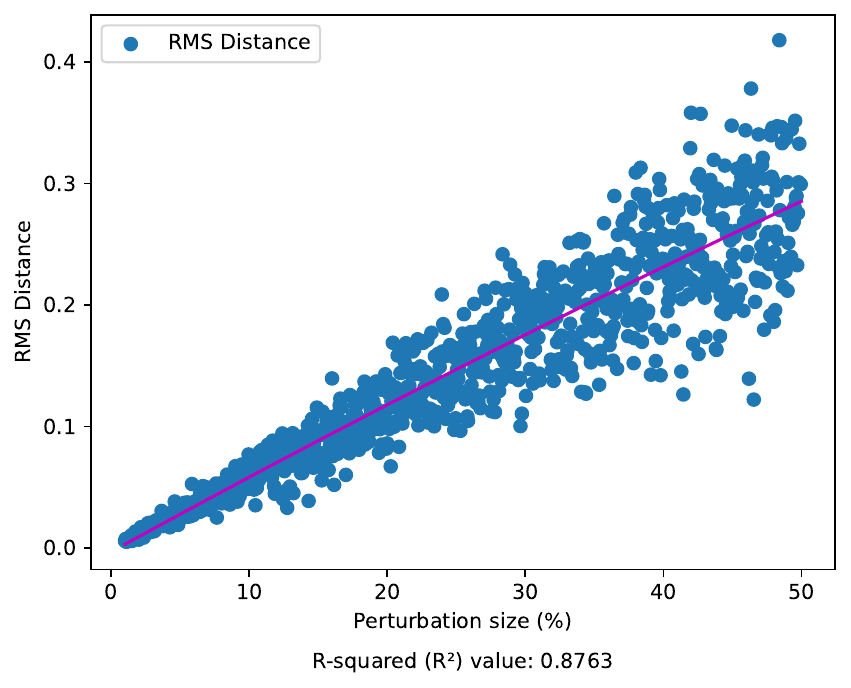}
        \caption{RMS Distance}
        \vspace{-3pt}
        \label{fig:Ni3S4_predict}
    \end{subfigure}
    \begin{subfigure}[t]{0.32\textwidth}
        \includegraphics[width=\textwidth]{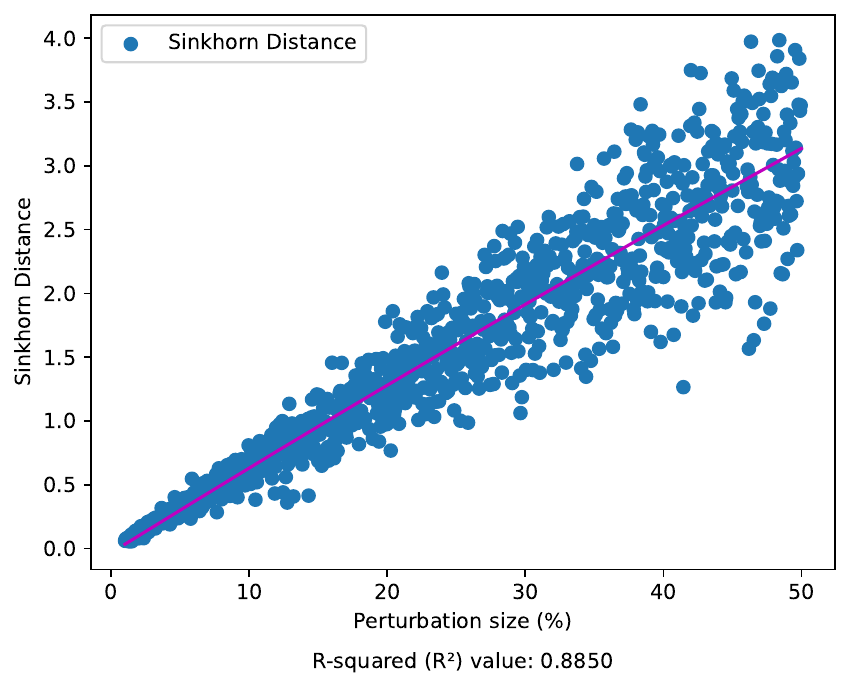}
        \caption{Sinkhorn Distance}
        \vspace{-3pt}
        \label{fig:NiS2_predict}
    \end{subfigure}\hfill
 \begin{subfigure}[t]{0.32\textwidth}
        \includegraphics[width=\textwidth]{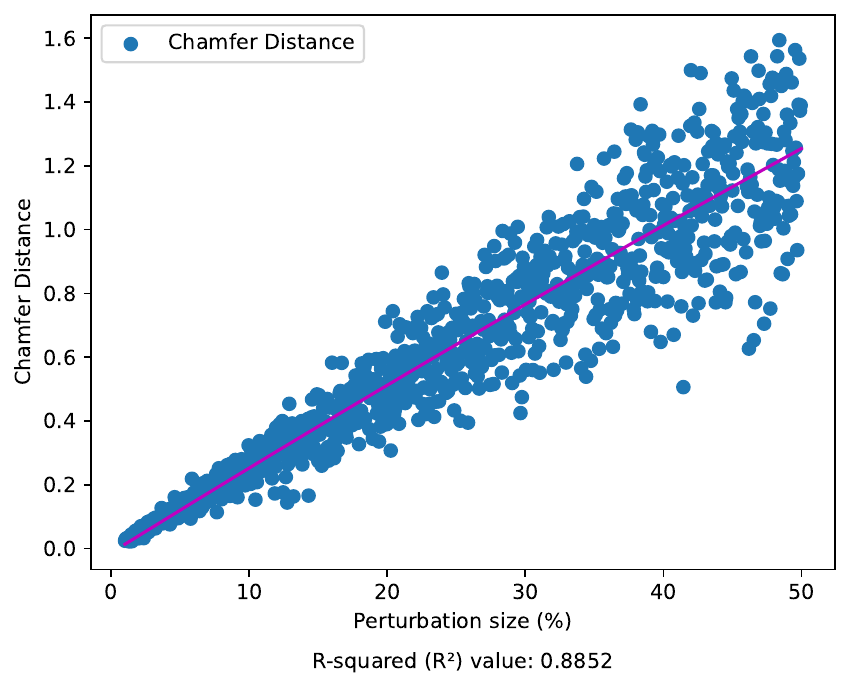}
        \caption{Chamfer Distance}
        \vspace{-3pt}
        \label{fig:GaBN2_target}
    \end{subfigure}
    \begin{subfigure}[t]{0.32\textwidth}
        \includegraphics[width=\textwidth]{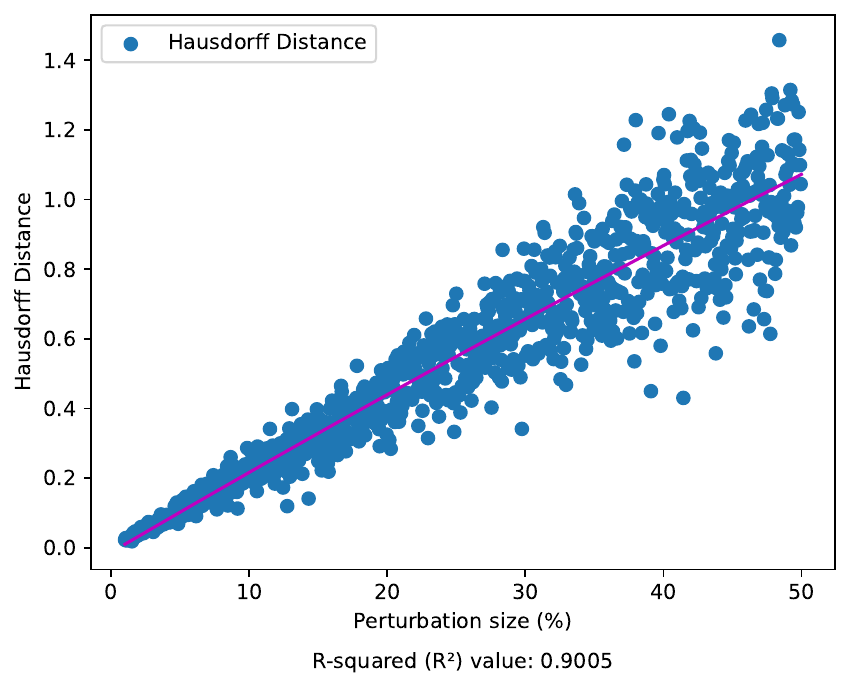}
        \caption{Hausdorff Distance}
        \vspace{-3pt}
        \label{fig:GaBN2_predict1}
    \end{subfigure}\hfill    
    \begin{subfigure}[t]{0.32\textwidth}
        \includegraphics[width=\textwidth]{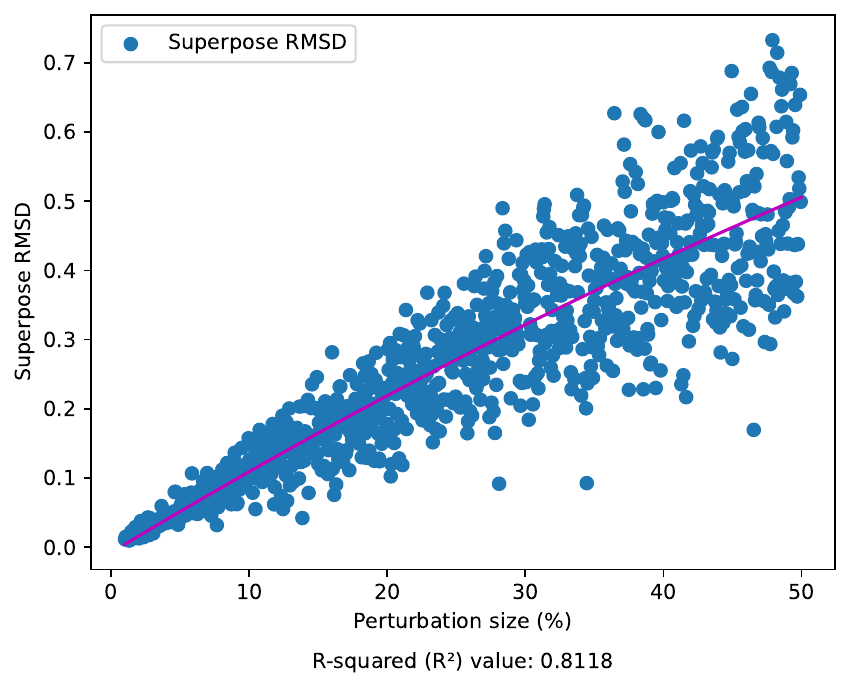}
        \caption{Superpose RMSD}
        \vspace{-3pt}
        \label{fig:GaBN2_predict2}
    \end{subfigure}\hfill    
    \begin{subfigure}[t]{0.32\textwidth}
        \includegraphics[width=\textwidth]{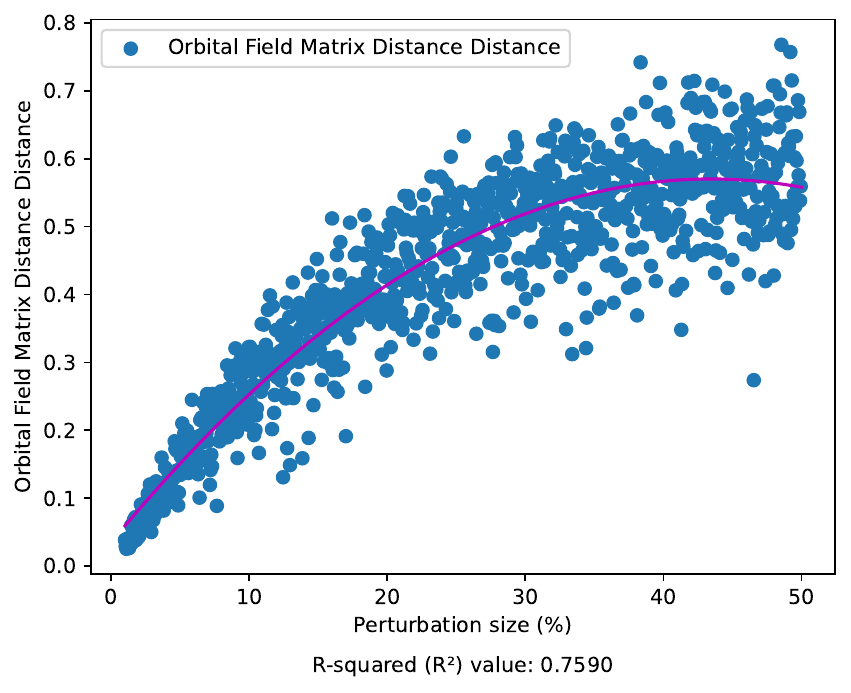}
        \caption{OFM Distance}
        \vspace{-3pt}
        \label{fig:SrTiO3_predict}
    \end{subfigure}\hfill
    \begin{subfigure}[t]{0.32\textwidth}
        \includegraphics[width=\textwidth]{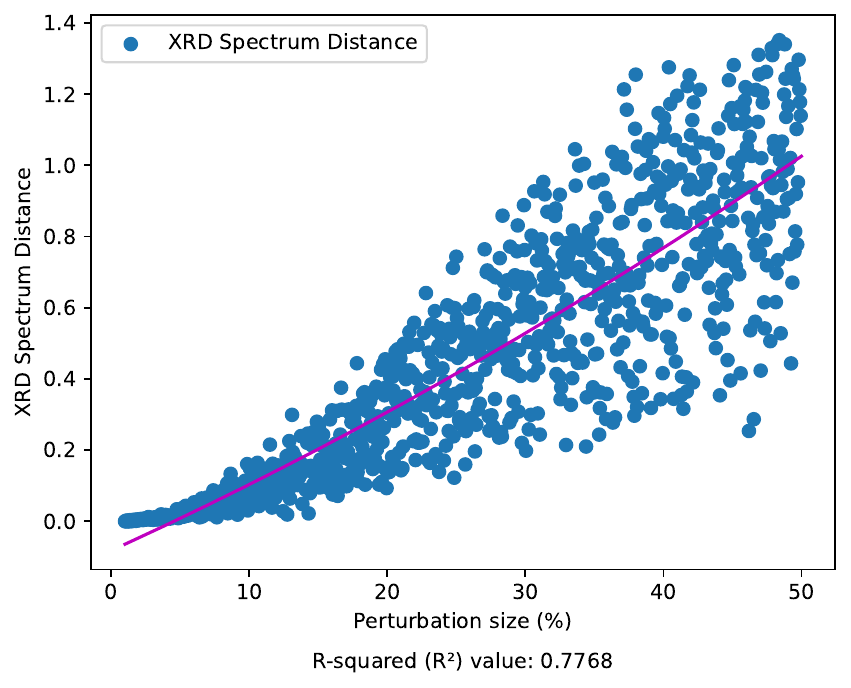}
        \caption{XRD Spectrum Distance}
        \vspace{-3pt}
        \label{fig:SrTiO3_predict}
    \end{subfigure}\hfill\textbf{}
        \begin{subfigure}[t]{0.32\textwidth}
        \includegraphics[width=\textwidth]{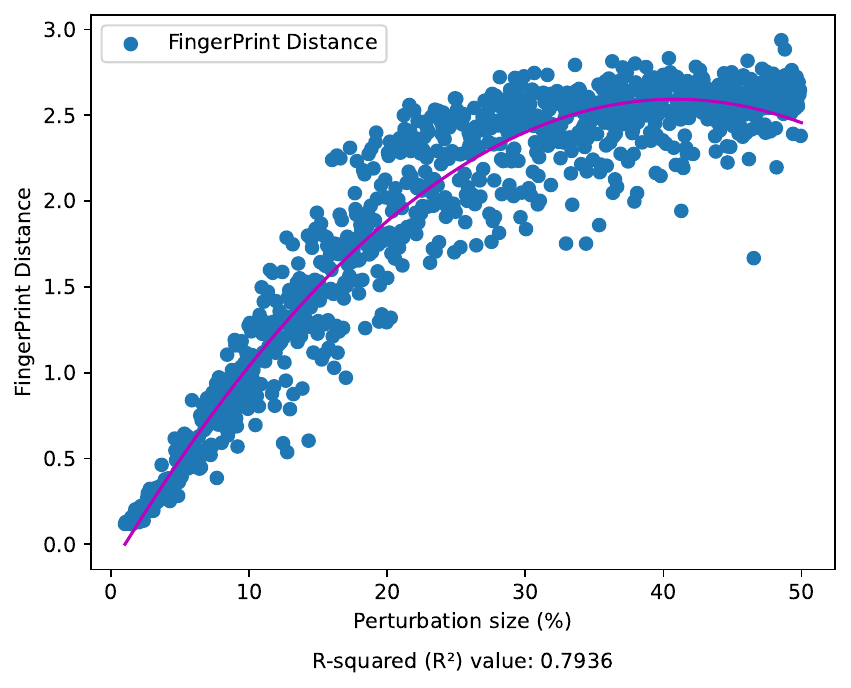}
        \caption{FingerPrint Distance}
        \vspace{-3pt}
        \label{fig:GaB3N4_predict1}
    \end{subfigure}\hfill  
   \caption{Structure distances vs perturbation size evaluated over the random perturbation structures of SrTiO\textsubscript{3}}
  \label{fig:perturb}
\end{figure}

Figure \ref{fig:perturb} (a) shows the parity plot of perturbation magnitude and the formation energy distances of the structures compared to the ground truth structure SrTiO$_3$. Here the formation energy is predicted using the universal machine learned M3GNet potential \cite{chen2022universal}. It can be found that as the perturbation percentage goes up, the energy difference between the perturbed structures and the stable structure also goes up. We can also find that the range of energy distances for a given perturbation percentage increases as the perturbation magnitude goes up, indicating the fact that highly disrupted structures tend to have diverse energy values. 
Figure \ref{fig:perturb} (b)-(l) shows the correlation between perturbation magnitude and eleven performance metrics. It has three types of correlations. Figure \ref{fig:perturb} (b)-(i) show the linear correlation of perturbation with respect to the following distance metrics, including Wyckoff RMSE, Wyckoff MAE, Anonymous RMS, RMS distance, Sinkhorn distance, Chamfer distance, Hausdorff Distance, and Superpose RMSD. Out of the eleven metrics, Wyckoff RMSE, RMS distance, and Hausdorff distance (Figure \ref{fig:perturb} (b)(e)(h)) show a higher degree of linearity. Relatively speaking, the remaining metrics demonstrate a certain degree of nonlinear correlation, including XRD Spectrum distance, OFM distance, and FingerPrint distance sorted by the degree of nonlinearity. All eleven metrics can be used to measure the similarity between candidate structures and the ground truth structure.

\begin{figure}[ht!] 
 \begin{subfigure}[t]{0.32\textwidth}
        \includegraphics[width=\textwidth]{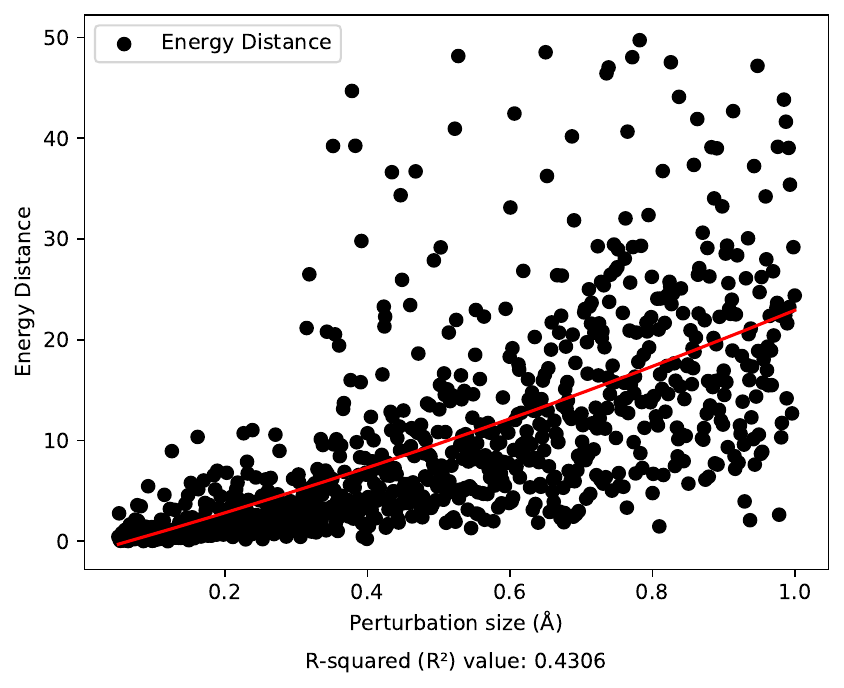}
        \caption{ZrSO (Energy Distance)}
        \vspace{-3pt}
        \label{fig:SrTiO3_target}
    \end{subfigure}
    \begin{subfigure}[t]{0.32\textwidth}
        \includegraphics[width=\textwidth]{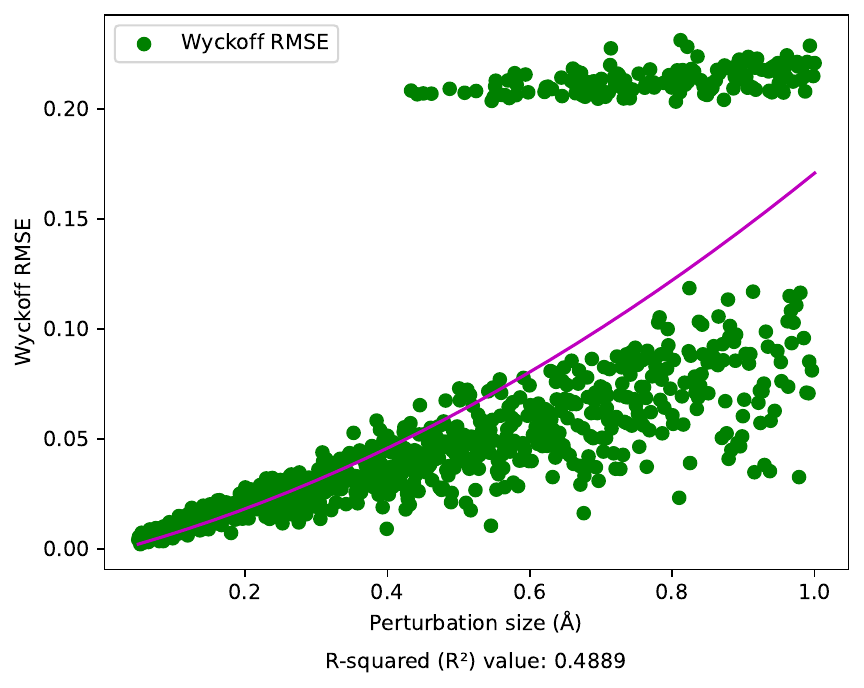}
        \caption{Wyckoff RMSE}
        \vspace{-3pt}
        \label{fig:Ni3S4_target}
    \end{subfigure}    
    \begin{subfigure}[t]{0.32\textwidth}
        \includegraphics[width=\textwidth]{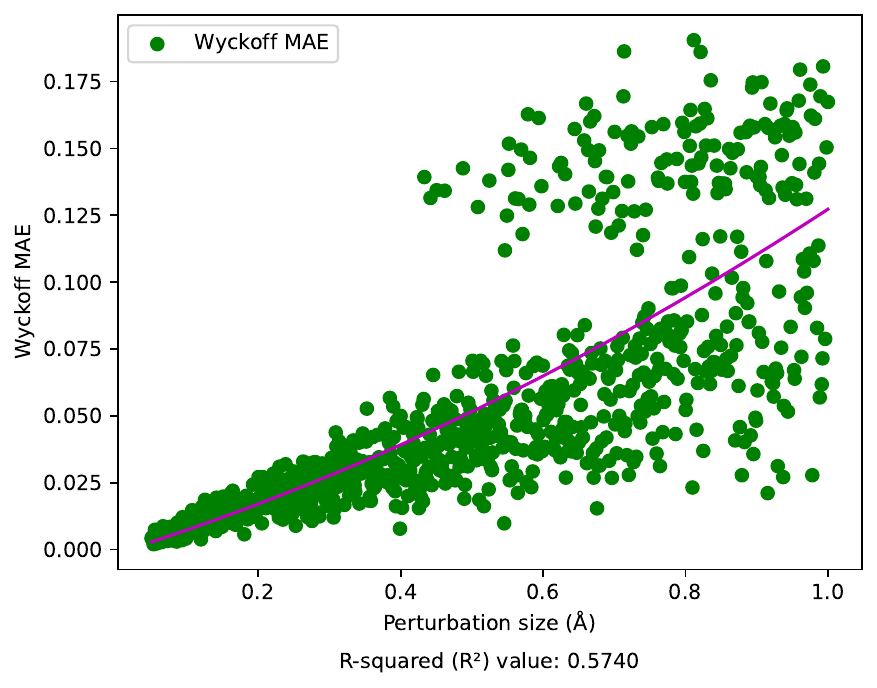}
        \caption{Wyckoff MAE}
        \vspace{-3pt}
        \label{fig:NiS2_target}
    \end{subfigure}    
    \begin{subfigure}[t]{0.32\textwidth}
        \includegraphics[width=\textwidth]{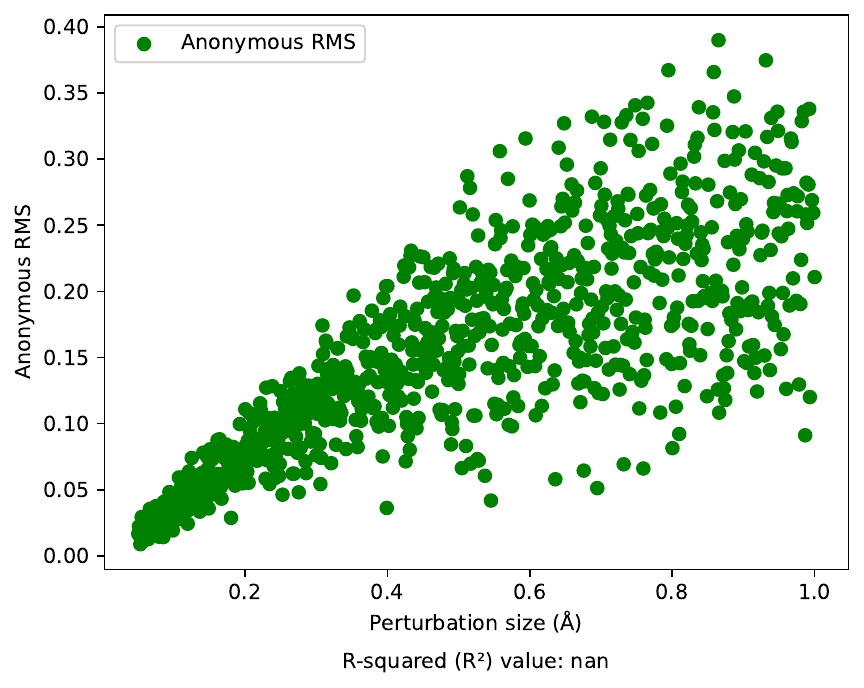}
        \caption{Anonymous RMS}
        \vspace{-3pt}
        \label{fig:Ni3S4_predict}
    \end{subfigure}
        \begin{subfigure}[t]{0.32\textwidth}
        \includegraphics[width=\textwidth]{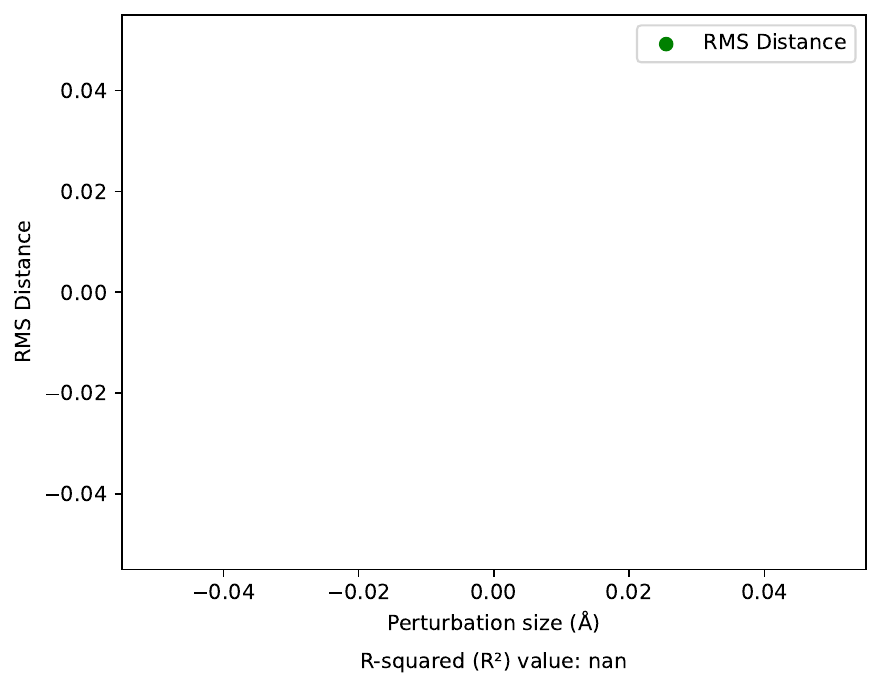}
        \caption{RMS Distance}
        \vspace{-3pt}
        \label{fig:Ni3S4_predict}
    \end{subfigure}
    \begin{subfigure}[t]{0.32\textwidth}
        \includegraphics[width=\textwidth]{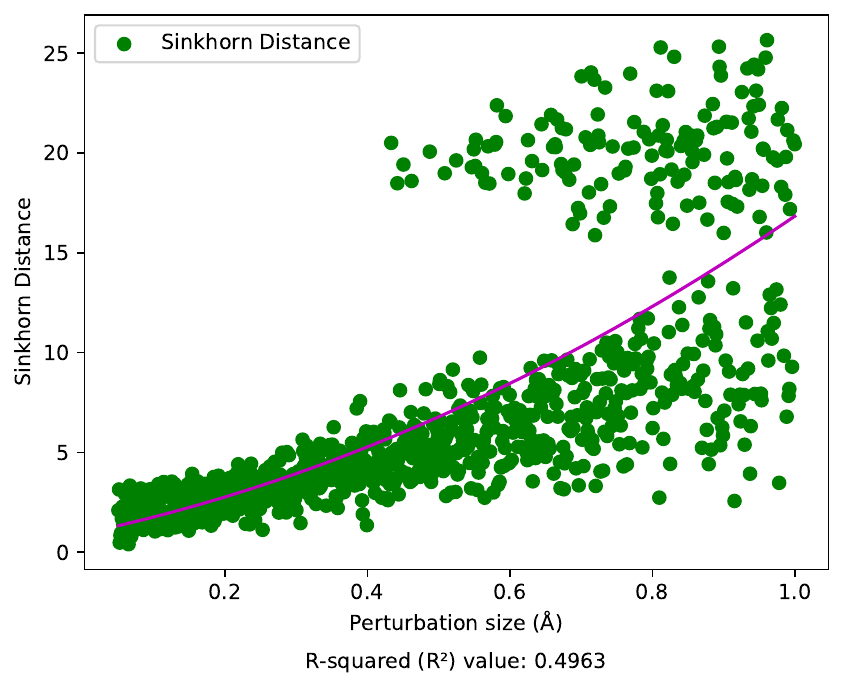}
        \caption{Sinkhorn Distance}
        \vspace{-3pt}
        \label{fig:NiS2_predict}
    \end{subfigure}\hfill
 \begin{subfigure}[t]{0.32\textwidth}
        \includegraphics[width=\textwidth]{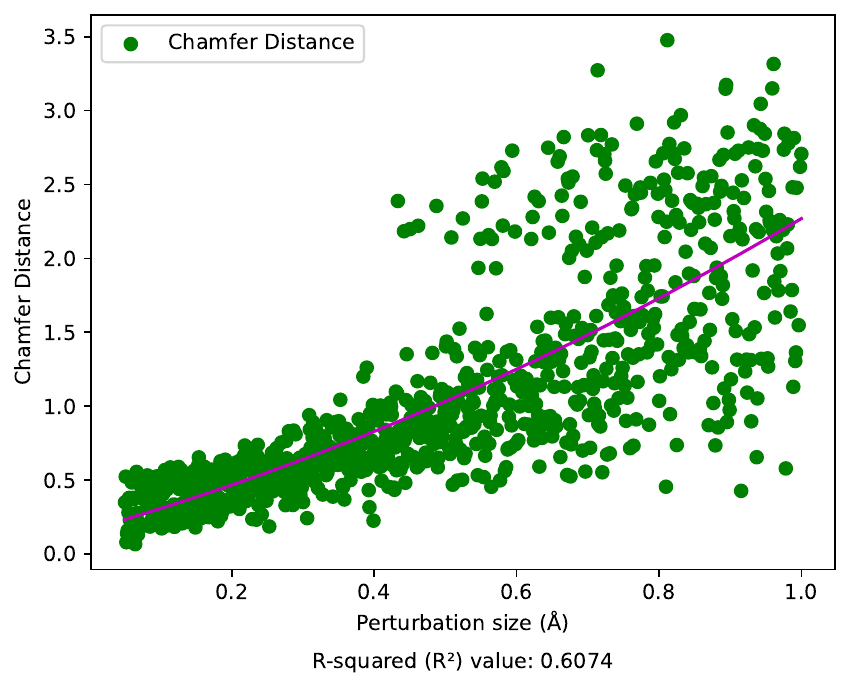}
        \caption{Chamfer Distance}
        \vspace{-3pt}
        \label{fig:GaBN2_target}
    \end{subfigure}
    \begin{subfigure}[t]{0.32\textwidth}
        \includegraphics[width=\textwidth]{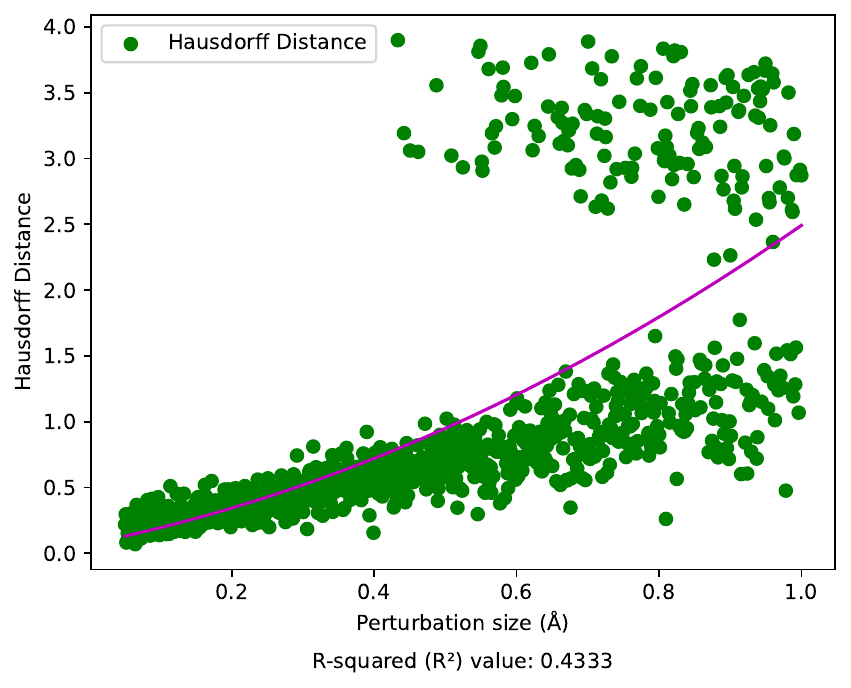}
        \caption{Hausdorff Distance}
        \vspace{-3pt}
        \label{fig:GaBN2_predict1}
    \end{subfigure}\hfill    
    \begin{subfigure}[t]{0.32\textwidth}
        \includegraphics[width=\textwidth]{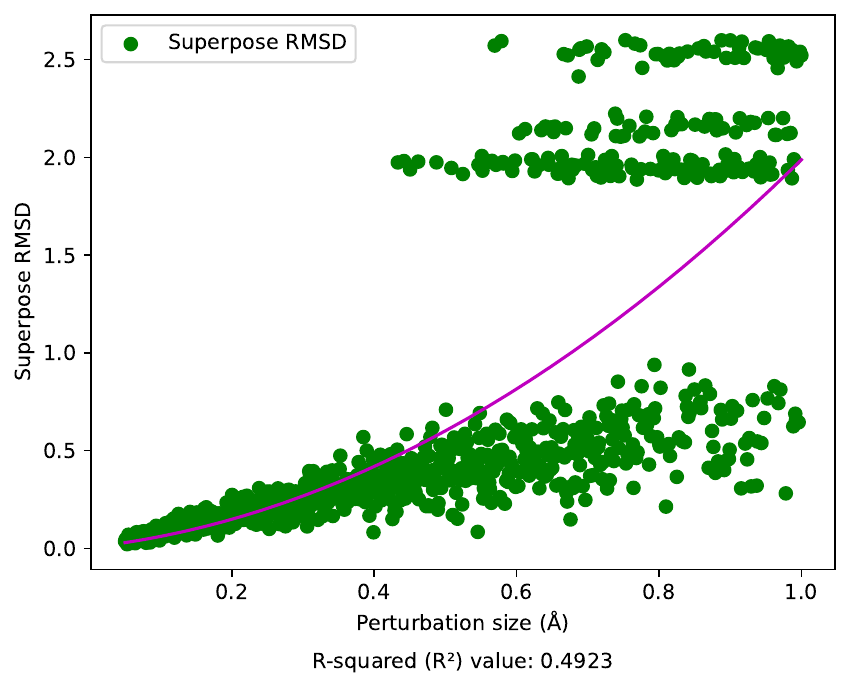}
        \caption{Superpose RMSD}
        \vspace{-3pt}
        \label{fig:GaBN2_predict2}
    \end{subfigure}\hfill
    \begin{subfigure}[t]{0.32\textwidth}
        \includegraphics[width=\textwidth]{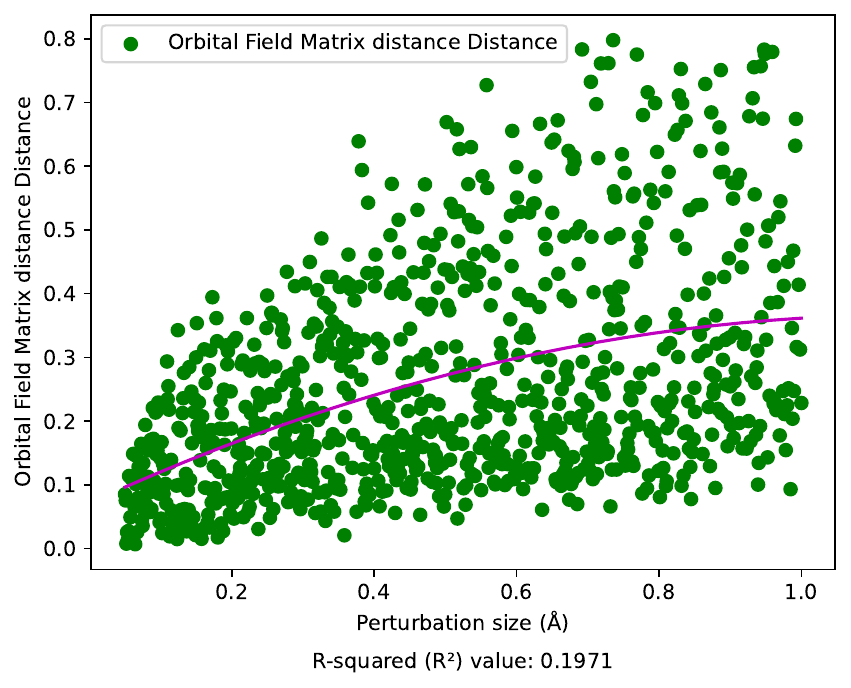}
        \caption{OFM Distance}
        \vspace{-3pt}
        \label{fig:SrTiO3_predict}
    \end{subfigure}\hfill
    \begin{subfigure}[t]{0.32\textwidth}
        \includegraphics[width=\textwidth]{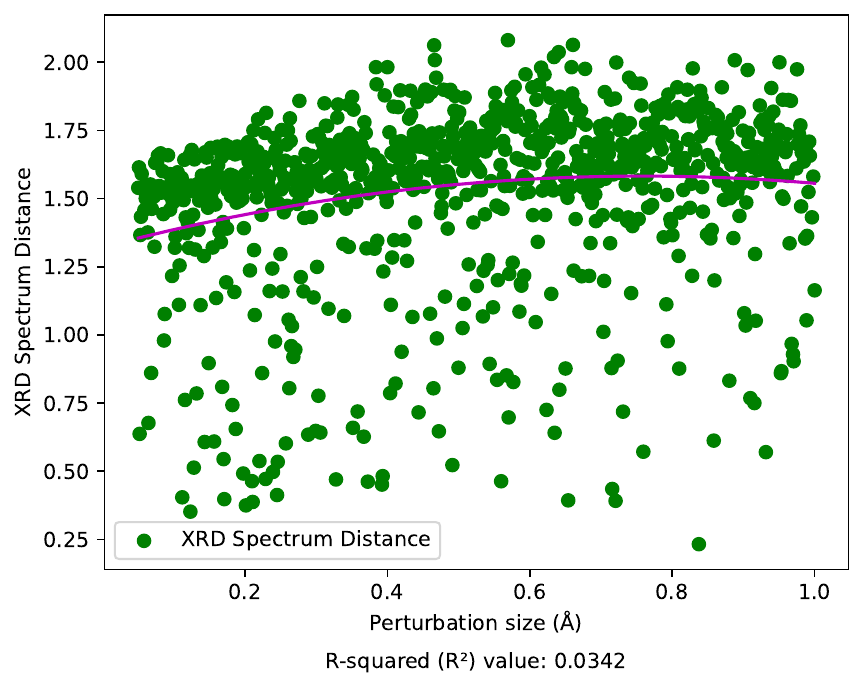}
        \caption{XRD Spectrum Distance}
        \vspace{-3pt}
        \label{fig:SrTiO3_predict}
    \end{subfigure}\hfill\textbf{}
        \begin{subfigure}[t]{0.32\textwidth}
        \includegraphics[width=\textwidth]{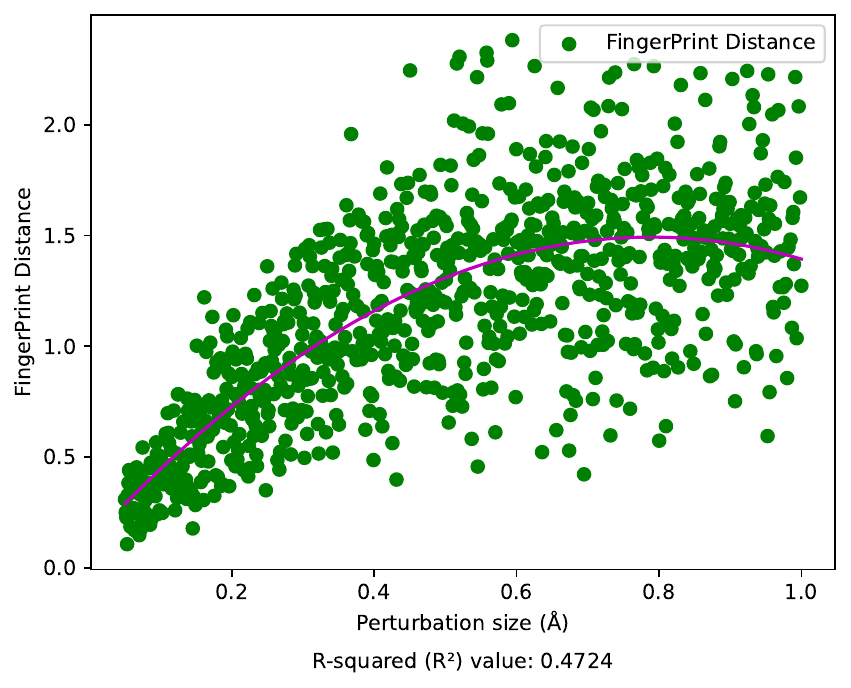}
        \caption{FingerPrint Distance}
        \vspace{-3pt}
        \label{fig:GaB3N4_predict1}
    \end{subfigure}\hfill  
   \caption{Structure distances vs perturbation size over the symmetrically perturbation structures of ZrSO. The dataset is generated by applying lattice a/b/c perturbation with small 5\% changes while fraction coordinates are perturbed with 2\% to 100\%. Space group remains unchanged.}
  \label{fig:perturbsymmetry}
\end{figure}

While the eleven performance metrics show a good correlation with the structure perturbation in Figure \ref{fig:perturb}, they are evaluated over the randomly perturbed structures which neglect the symmetry relationships among equivalent Wyckoff atomic sites. However, many efficient CSP algorithms use symmetry-obeying search operators which do not violate the atomic symmetry relationship during the coordinate search. To simulate this situation, we generate a second set of symmetry-preserving perturbed structures from the ground truth structure ZrSO. Figure \ref{fig:perturbsymmetry} shows the correlations of performance metrics with respect to the perturbation magnitude.

\FloatBarrier

Compared to the random perturbation structures of SrTiO$_3$, symmetrical perturbations, in which the perturbation is applied to the coordinates of Wyckoff sites, may lead to more pronounced structural changes. It can be found that the correlation between perturbation magnitude and performance metrics is weaker (Figure \ref{fig:perturbsymmetry}). Because symmetrical structures possess internal repetition and symmetry, and a perturbation in one point can propagate equivalent perturbations in other points. Consequently, accurately describing the perturbation based solely on its magnitude becomes challenging. In such scenarios, the influence of perturbations can extend beyond their immediate vicinity, impacting a broader range of elements or variables within the system. Therefore, to fully comprehend and analyze the effects of symmetrical perturbations, it is crucial to possess a comprehensive understanding of the system's dynamics and interactions. From Figure \ref{fig:perturbsymmetry} (a), we can find that the perturbations quickly lead to high variations of the energies of the perturbed structures despite that the pattern is similar to random perturbation when the perturbation magnitude/percentage is small. For the distance metrics in Figure \ref{fig:perturbsymmetry} (b, c, f, g, h, i), we all observe regular patterns at the bottom which are similar to the trends in the random perturbation results (Figure \ref{fig:perturb}). The top right overhead dots show the impact of symmetric perturbation: a relatively small perturbation can also cause big distance changes. We also find that Figures \ref{fig:perturbsymmetry}(j, k, l) have much higher variation than those in Figures \ref{fig:perturb}(j, k, l) respectively.  These results show that our selected distance metrics tend to have higher variation when used to evaluate the structure similarity for symmetrically perturbed structures. It is also found that the RMS distance is not even usable for symmetric perturbation (Figure \ref{fig:perturbsymmetry} (e)).

\subsection{Using distance metrics to compare CSP algorithms}
From the Material Project Database \cite{jain2013commentary}, we individually choose five crystal structures for each type of composition encompassing binary, ternary, and quaternary, which generates a total of fifteen test structures. We also added CaS (mp-1672) as one additional target. To compare the performance of three GN-OA algorithms \cite{cheng2022crystal} based on three optimization methods including random searching (RAS), Bayesian optimization (BO), and particle swarm optimization (PSO), we applied these three algorithms (GN-OA-RAS, GN-OA-BO, GN-OA-PSO) to the 16 targets. For RAS and BO, we set the init population size to 200 and the total number of iterations to 20,000. For the PSO algorithm, we set the init population size to be 200 and the generation number is 100.

Table \ref{tab:CaS} shows the distance metrics between the ground truth structure of CaS and its predicted structures by three CSP algorithms, two of which (GN-OA-RAS and GN-OA-BO) successfully predict the ground-state structures within 5000 iteration steps. The computed distance metrics demonstrate similar performance across all measures. The energy distances for RAS, BO, and PSO are 3.4074, 2.9220, and 2.9493, respectively. Additionally, the XRD spectrum distance values of 2.8915, 2.8651, and 2.8636, along with the corresponding OFM distance values of 0.1535, 0.1419, and 0.1425 for RAS, BO, and PSO, indicate highly consistent results. Notably, both the edit graph distance and the fingerprint distance exhibit perfect matches with the ground truth with the values of 0 for each algorithm. Figure \ref{fig:CaS} shows the comparison of the ground truth with the predicted crystal structures of CaS by the RAS, BO, and PSO algorithms. Figure \ref{fig:CaS} (b), figure \ref{fig:CaS} (c), and figure \ref{fig:CaS} (d) exhibit a striking similarity in structure to Figure \ref{fig:CaS} (a). 

Table \ref{tab:ScBe5} shows the distance metrics of the predicted versus the ground truth for the binary target structure ScBe$_5$. First, we found that the PSO algorithm achieved the highest performance for all distances except the formation energy. The energy distances are 12.4945, 23.6137, and 4.7720 for RAS, BO, and PSO, respectively. Figure \ref{fig:ScBe5} (d) demonstrates a more symmetrical and similar structure compared to Figures \ref{fig:ScBe5}(b) and \ref{fig:ScBe5} (c), which implies that Figure \ref{fig:ScBe5} (d) likely exhibits a balanced distribution of elements and a logical arrangement that serves its purpose effectively. The Sinkhorn distance, the Chamfer distance, and the Hausdorff distance of PSO are also significantly smaller than those of RAS and BO, which are 2.7904, 0.9301, and 1.2743, respectively. All the distances indicate a higher similarity between the structure presented in Figure \ref{fig:ScBe5} (d) and the ground truth structure. As shown in Table \ref{tab:LiZrO2}, the distances to the ground structure for both BO and PSO results are nearly identical, with energy distances of 5.6953 and 5.6967, respectively, compared to 68.7212 for RAS. Although the formation energy does not differ significantly among RAS, BO, and PSO, the structure of BO and PSO in Figure \ref{fig:LiZrO2} (c) and Figure \ref{fig:LiZrO2} (d) shows a more similar structure because they have almost the same values of formation energy. Figure \ref{fig:LiZrO2} (b) shows a higher symmetry compared to Figure \ref{fig:LiZrO2} (c) and Figure \ref{fig:LiZrO2} (d), which may also be the reason why the Sinkhorn distance, the Chamfer distance, and the Hausdorff distances indicate better performance for RAS. For the quaternary material LiTiSe$_2$O (Table \ref{tab:LiTiSe2O}), the BO algorithm achieves better performance in terms of most distance metrics. Among all the distance metrics, BO shows the best performance for the energy distance with a value of 19.2273 compared to 93.8576 and 106.6080 of RAS and PSO respectively. In addition, BO outperforms RAS and PSO in terms of the more challenging indicators including Wyckoff RMSE and Wyckoff MAE, with values of 0.3464 and 0.2458, respectively. Similarly, it has the best results in terms of the Sinkhorn distance, the Chamfer distance, and the Hausdorff distance metrics which are 23.6450, 2.8741, and 2.6054. Figure \ref{fig:LiTiSe2O} (c) shows the predicted structure of BO, which has a more symmetrical structure compared to Figure \ref{fig:LiTiSe2O} (b) and Figure \ref{fig:LiTiSe2O} (d) predicted by RAS and PSO. The figures and tables presented above effectively demonstrate that our distance metrics accurately capture the differences between crystal structures with the ground truths, highlighting that more symmetrical and stable structures tend to allow the CSP algorithms to achieve better performance. Performance evaluation metrics of additional the 12 targets are shown in the Supplementary file Table S1 to S12.

\begin{table}[htbp]
  \centering
  \begin{minipage}{0.5\textwidth}
    \centering
    \caption{A metrics table generated by comparing the ground truth structure of CaS with the structures obtained using three different optimization algorithms from GN-OA: Random Acceleration Search (RAS), Particle Swarm Optimization (PSO), and Bayesian Optimization (BO).}
    \begin{tabular}{|llll|}
    
      \hline
\multicolumn{4}{|c|}{\textbf{CaS}}                                                                                                                                     \\ \hline
\multicolumn{1}{|l|}{Algorithm}                                     & \multicolumn{1}{l|}{RAS}              & \multicolumn{1}{l|}{BO}               & PSO              \\ \hline
\multicolumn{1}{|l|}{Formation Energy}                              & \multicolumn{1}{l|}{\textbf{-1.4944}} & \multicolumn{1}{l|}{-1.4955} & -1.4955 \\ \hline
\multicolumn{1}{|l|}{\cellcolor[HTML]{FFFFFF}Energy Distance}       & \multicolumn{1}{l|}{3.4074}           & \multicolumn{1}{l|}{\textbf{2.9220}}  & 2.9493           \\ \hline
\multicolumn{1}{|l|}{\cellcolor[HTML]{FFFFFF}Wyckoff RMSE}          & \multicolumn{1}{l|}{N/A}              & \multicolumn{1}{l|}{N/A}              & N/A              \\ \hline
\multicolumn{1}{|l|}{\cellcolor[HTML]{FFFFFF}Wyckoff MAE}           & \multicolumn{1}{l|}{N/A}              & \multicolumn{1}{l|}{N/A}              & N/A              \\ \hline
\multicolumn{1}{|l|}{\cellcolor[HTML]{FFFFFF}Anonymous RMS}         & \multicolumn{1}{l|}{N/A}              & \multicolumn{1}{l|}{N/A}              & N/A              \\ \hline
\multicolumn{1}{|l|}{\cellcolor[HTML]{FFFFFF}RMS Distance}          & \multicolumn{1}{l|}{N/A}              & \multicolumn{1}{l|}{N/A}              & N/A              \\ \hline
\multicolumn{1}{|l|}{\cellcolor[HTML]{FFFFFF}Sinkhorn Distance}     & \multicolumn{1}{l|}{3.2478}           & \multicolumn{1}{l|}{\textbf{2.9730}}   & 2.9888           \\ \hline
\multicolumn{1}{|l|}{\cellcolor[HTML]{FFFFFF}Chamfer Distance}      & \multicolumn{1}{l|}{\textbf{0.8120}}   & \multicolumn{1}{l|}{0.7432}           & 0.7472           \\ \hline
\multicolumn{1}{|l|}{\cellcolor[HTML]{FFFFFF}Hausdorff Distance}    & \multicolumn{1}{l|}{0.6268}           & \multicolumn{1}{l|}{\textbf{0.5738}}  & 0.5768           \\ \hline
\multicolumn{1}{|l|}{\cellcolor[HTML]{FFFFFF}Superpose RMSD}        & \multicolumn{1}{l|}{\textbf{1.8412}}           & \multicolumn{1}{l|}{2.3290}            & \textbf{1.8412}  \\ \hline
\multicolumn{1}{|l|}{Edit Graph Distance}                           & \multicolumn{1}{l|}{\textbf{0}}       & \multicolumn{1}{l|}{\textbf{0}}       & \textbf{0}       \\ \hline
\multicolumn{1}{|l|}{\cellcolor[HTML]{FFFFFF}FingerPrint Distance}  & \multicolumn{1}{l|}{\textbf{0}}       & \multicolumn{1}{l|}{\textbf{0}}       & \textbf{0}       \\ \hline
\multicolumn{1}{|l|}{\cellcolor[HTML]{FFFFFF}XRD Spectrum Distance} & \multicolumn{1}{l|}{2.8915}           & \multicolumn{1}{l|}{2.8651}           & \textbf{2.8636}  \\ \hline
\multicolumn{1}{|l|}{\cellcolor[HTML]{FFFFFF}OFM Distance}          & \multicolumn{1}{l|}{0.1535}           & \multicolumn{1}{l|}{\textbf{0.1419}}  & 0.1425           \\ \hline
\end{tabular}

\label{tab:CaS}
  \end{minipage}%
  \hfill
  \begin{minipage}{0.4\textwidth}
    \centering
    \begin{subfigure}[b]{0.4\textwidth}
      \includegraphics[width=\textwidth]{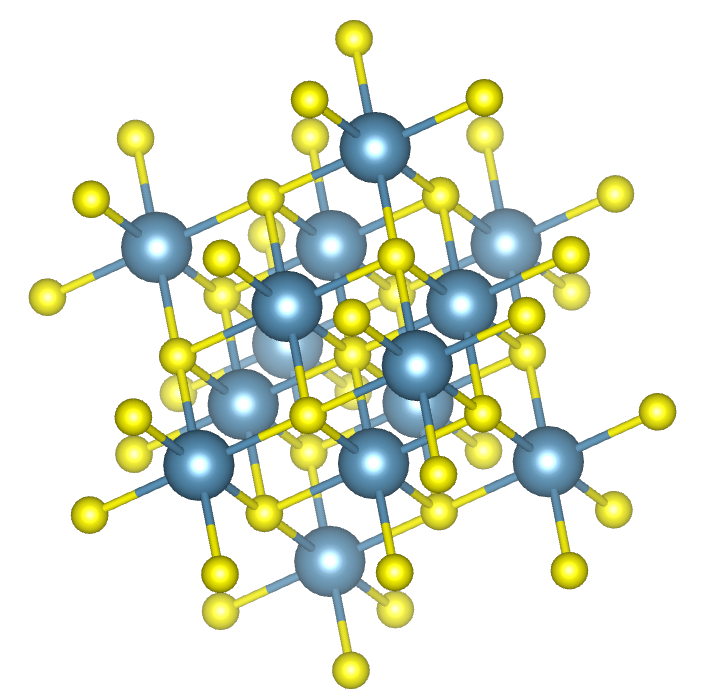}
      \caption{Ground truth}
      \label{fig:CaS-a}
    \end{subfigure}\hfill
    \begin{subfigure}[b]{0.45\textwidth}
      \includegraphics[width=\textwidth]{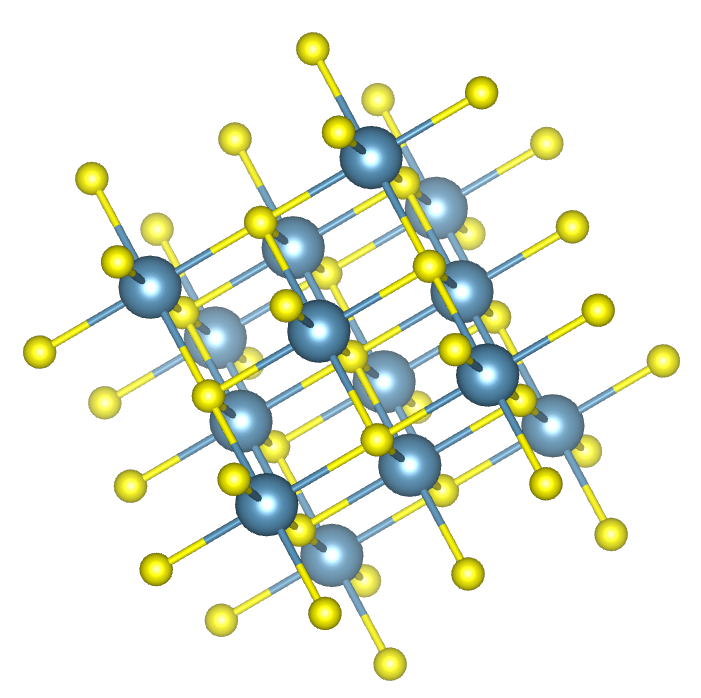}
      \caption{Predicted by RAS}
      \label{fig:CaS-b}
    \end{subfigure}\\[0.45em]
    \begin{subfigure}[b]{0.45\textwidth}
      \includegraphics[width=\textwidth]{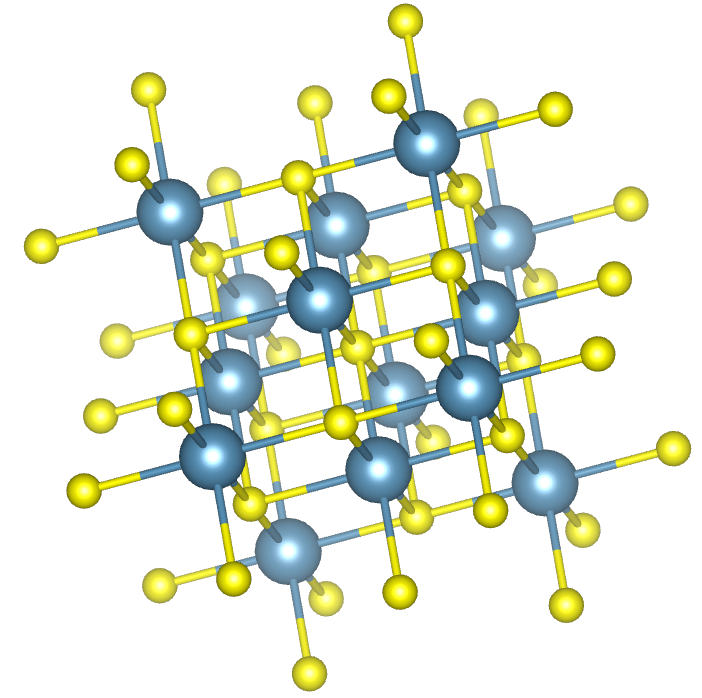}
      \caption{Predicted by BO}
      \label{fig:CaS-c}
    \end{subfigure}\hfill
    \begin{subfigure}[b]{0.45\textwidth}
      \includegraphics[width=\textwidth]{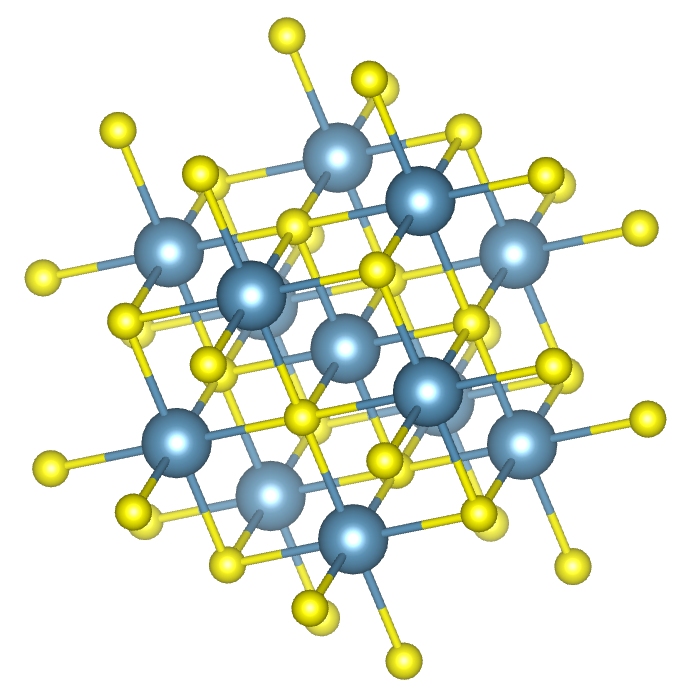}
      \caption{Predicted by PSO}
      \label{fig:CaS-d}
    \end{subfigure}
    \captionof{figure}{Comparison of the ground truth and predicted crystal structures of CaS by the RAS, BO, and PSO algorithms.}
    \label{fig:CaS}
  \end{minipage}
\end{table}

\begin{table}[htbp]
  \centering
  \begin{minipage}{0.5\textwidth}
    \centering
    \caption{A metric table generated by comparing the ground truth structure of ScBe$_5$ with the structures obtained using three different optimization algorithms from GN-OA: Random Acceleration Search (RAS), Particle Swarm Optimization (PSO), and Bayesian Optimization (BO).}
    \begin{tabular}{|llll|}
      \hline
\multicolumn{4}{|c|}{\textbf{ScBe$_5$}}                                                                                                                                \\ \hline
\multicolumn{1}{|l|}{Algorithm}                                     & \multicolumn{1}{l|}{RAS}               & \multicolumn{1}{l|}{BO}                & PSO               \\ \hline
\multicolumn{1}{|l|}{Formation Energy}                              & \multicolumn{1}{l|}{0.0178} & \multicolumn{1}{l|}{\textbf{-0.3356}} & -0.2931 \\ \hline
\multicolumn{1}{|l|}{\cellcolor[HTML]{FFFFFF}Energy Distance}        & \multicolumn{1}{l|}{12.4945}  & \multicolumn{1}{l|}{23.6137}           & \textbf{4.7720}    \\ \hline
\multicolumn{1}{|l|}{\cellcolor[HTML]{FFFFFF}Wyckoff RMSE}          & \multicolumn{1}{l|}{N/A}     & \multicolumn{1}{l|}{N/A}              & N/A              \\ \hline
\multicolumn{1}{|l|}{\cellcolor[HTML]{FFFFFF}Wyckoff MAE}           & \multicolumn{1}{l|}{N/A}     & \multicolumn{1}{l|}{N/A}              & N/A              \\ \hline
\multicolumn{1}{|l|}{\cellcolor[HTML]{FFFFFF}Anonymous RMS}         & \multicolumn{1}{l|}{N/A}     & \multicolumn{1}{l|}{N/A}              & \textbf{0.4535}            \\ \hline
\multicolumn{1}{|l|}{\cellcolor[HTML]{FFFFFF}RMS Distance}          & \multicolumn{1}{l|}{N/A}     & \multicolumn{1}{l|}{N/A}              & \textbf{0.4535}   \\ \hline
\multicolumn{1}{|l|}{\cellcolor[HTML]{FFFFFF}Sinkhorn Distance}     & \multicolumn{1}{l|}{45.1382}  & \multicolumn{1}{l|}{47.2978}           & \textbf{2.7904}   \\ \hline
\multicolumn{1}{|l|}{\cellcolor[HTML]{FFFFFF}Chamfer Distance}      & \multicolumn{1}{l|}{9.3233}   & \multicolumn{1}{l|}{10.9360}            & \textbf{0.9301}   \\ \hline
\multicolumn{1}{|l|}{\cellcolor[HTML]{FFFFFF}Hausdorff Distance}    & \multicolumn{1}{l|}{13.9499}  & \multicolumn{1}{l|}{11.3823}           & \textbf{1.2743}   \\ \hline
\multicolumn{1}{|l|}{\cellcolor[HTML]{FFFFFF}Superpose RMSD}        & \multicolumn{1}{l|}{1.6628}   & \multicolumn{1}{l|}{1.6343}            & \textbf{1.6952}   \\ \hline
\multicolumn{1}{|l|}{Edit Graph Distance}                           & \multicolumn{1}{l|}{7}       & \multicolumn{1}{l|}{8}       & \textbf{3}       \\ \hline
\multicolumn{1}{|l|}{\cellcolor[HTML]{FFFFFF}FingerPrint Distance}  & \multicolumn{1}{l|}{2.3392}       & \multicolumn{1}{l|}{2.4184}     & \textbf{1.7783}       \\ \hline
\multicolumn{1}{|l|}{\cellcolor[HTML]{FFFFFF}XRD Spectrum Distance} & \multicolumn{1}{l|}{2.2394}           & \multicolumn{1}{l|}{1.9878}           & \textbf{1.6491}  \\ \hline
\multicolumn{1}{|l|}{\cellcolor[HTML]{FFFFFF}OFM Distance}          & \multicolumn{1}{l|}{0.9898}           & \multicolumn{1}{l|}{1.1039}  & \textbf{0.4494}           \\ \hline
\end{tabular}

\label{tab:ScBe5}
  \end{minipage}%
  \hfill
  \begin{minipage}{0.4\textwidth}
    \centering
    \begin{subfigure}[b]{0.4\textwidth}
      \includegraphics[width=\textwidth]{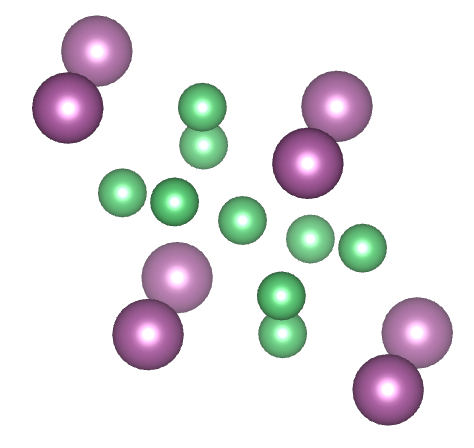}
      \caption{Ground truth}
      \label{fig:ScBe5-a}
    \end{subfigure}\hfill
    \begin{subfigure}[b]{0.45\textwidth}
      \includegraphics[width=\textwidth]{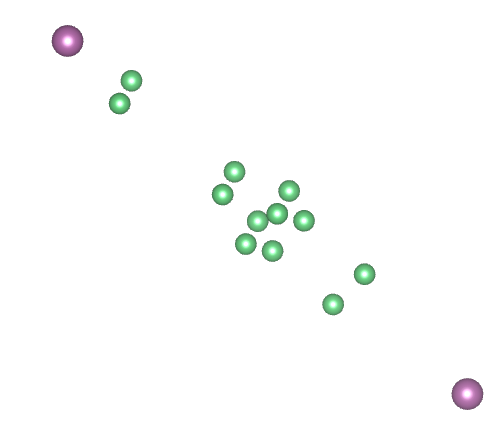}
      \caption{Predicted by RAS}
      \label{fig:ScBe5-b}
    \end{subfigure}\\[0.45em]
    \begin{subfigure}[b]{0.45\textwidth}
      \includegraphics[width=\textwidth]{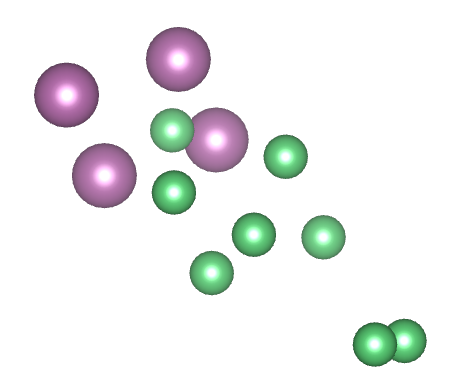}
      \caption{Predicted by BO}
      \label{fig:ScBe5-c}
    \end{subfigure}\hfill
    \begin{subfigure}[b]{0.45\textwidth}
      \includegraphics[width=\textwidth]{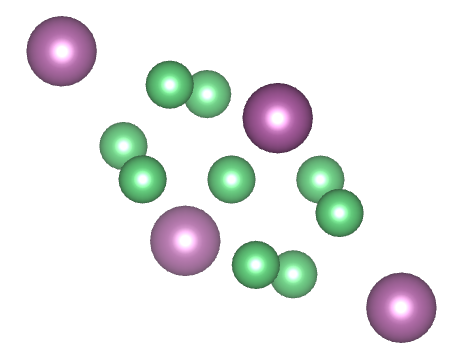}
      \caption{Predicted by  PSO}
      \label{fig:ScBe5-d}
    \end{subfigure}
    \captionof{figure}{Comparison of the ground truth and predicted crystal structures of ScBe$_5$ by the RAS, BO, and PSO algorithms.}
    \label{fig:ScBe5}
  \end{minipage}
\end{table}

\begin{table}[htbp]
  \centering
  \begin{minipage}{0.5\textwidth}
    \centering
    \caption{A metrics table generated by comparing the ground truth structure of LiZrO$_2$ with the structures obtained using three different optimization algorithms from GN-OA: Random Acceleration Search (RAS), Particle Swarm Optimization (PSO), and Bayesian Optimization (BO).}
    \begin{tabular}{|llll|}
\hline
\multicolumn{4}{|c|}{\textbf{LiZrO$_2$}}                                                                                                                                           \\ \hline
\multicolumn{1}{|l|}{Algorithm}                                     & \multicolumn{1}{l|}{RAS}              & \multicolumn{1}{l|}{BO}               & PSO              \\ \hline
\multicolumn{1}{|l|}{Formation Energy}                              & \multicolumn{1}{l|}{-1.3519} & \multicolumn{1}{l|}{\textbf{-1.4883}} & \textbf{-1.4883} \\ \hline
\multicolumn{1}{|l|}{\cellcolor[HTML]{FFFFFF}Energy Distance}        & \multicolumn{1}{l|}{68.7212} & \multicolumn{1}{l|}{\textbf{5.6953}}  & 5.6967  \\ \hline
\multicolumn{1}{|l|}{\cellcolor[HTML]{FFFFFF}Wyckoff RMSE}          & \multicolumn{1}{l|}{N/A}             & \multicolumn{1}{l|}{N/A}             & N/A             \\ \hline
\multicolumn{1}{|l|}{\cellcolor[HTML]{FFFFFF}Wyckoff MAE}           & \multicolumn{1}{l|}{N/A}             & \multicolumn{1}{l|}{N/A}             & N/A             \\ \hline
\multicolumn{1}{|l|}{\cellcolor[HTML]{FFFFFF}Anonymous RMS}         & \multicolumn{1}{l|}{N/A}             & \multicolumn{1}{l|}{N/A}             & N/A             \\ \hline
\multicolumn{1}{|l|}{\cellcolor[HTML]{FFFFFF}RMS Distance}          & \multicolumn{1}{l|}{N/A}             & \multicolumn{1}{l|}{N/A}             & N/A    \\ \hline
\multicolumn{1}{|l|}{\cellcolor[HTML]{FFFFFF}Sinkhorn Distance}     & \multicolumn{1}{l|}{\textbf{31.7423}} & \multicolumn{1}{l|}{37.2488}          & 37.0778 \\ \hline
\multicolumn{1}{|l|}{\cellcolor[HTML]{FFFFFF}Chamfer Distance}      & \multicolumn{1}{l|}{\textbf{3.0679}}  & \multicolumn{1}{l|}{3.5888}           & 3.6062  \\ \hline
\multicolumn{1}{|l|}{\cellcolor[HTML]{FFFFFF}Hausdorff Distance}    & \multicolumn{1}{l|}{\textbf{3.5578}}  & \multicolumn{1}{l|}{4.4537}           & 4.6748  \\ \hline
\multicolumn{1}{|l|}{\cellcolor[HTML]{FFFFFF}Superpose RMSD}        & \multicolumn{1}{l|}{2.9837}           & \multicolumn{1}{l|}{\textbf{2.8446}}  & 2.9019  \\ \hline
\multicolumn{1}{|l|}{Edit Graph Distance}                           & \multicolumn{1}{l|}{\textbf{33}} & \multicolumn{1}{l|}{36}          & 36 \\ \hline
\multicolumn{1}{|l|}{\cellcolor[HTML]{FFFFFF}FingerPrint Distance}  & \multicolumn{1}{l|}{3.5050}           & \multicolumn{1}{l|}{\textbf{2.8482}}  & \textbf{2.8482}  \\ \hline
\multicolumn{1}{|l|}{\cellcolor[HTML]{FFFFFF}XRD Spectrum Distance} & \multicolumn{1}{l|}{\textbf{1.7977}}  & \multicolumn{1}{l|}{2.0243}  & 2.0243  \\ \hline
\multicolumn{1}{|l|}{\cellcolor[HTML]{FFFFFF}OFM Distance}          & \multicolumn{1}{l|}{0.4376}           & \multicolumn{1}{l|}{\textbf{0.2690}}  & 0.2691  \\ \hline
\end{tabular}

\label{tab:LiZrO2}
  \end{minipage}%
  \hfill
  \begin{minipage}{0.45\textwidth}
    \centering
    \begin{subfigure}[b]{0.45\textwidth}
      \includegraphics[width=\textwidth]{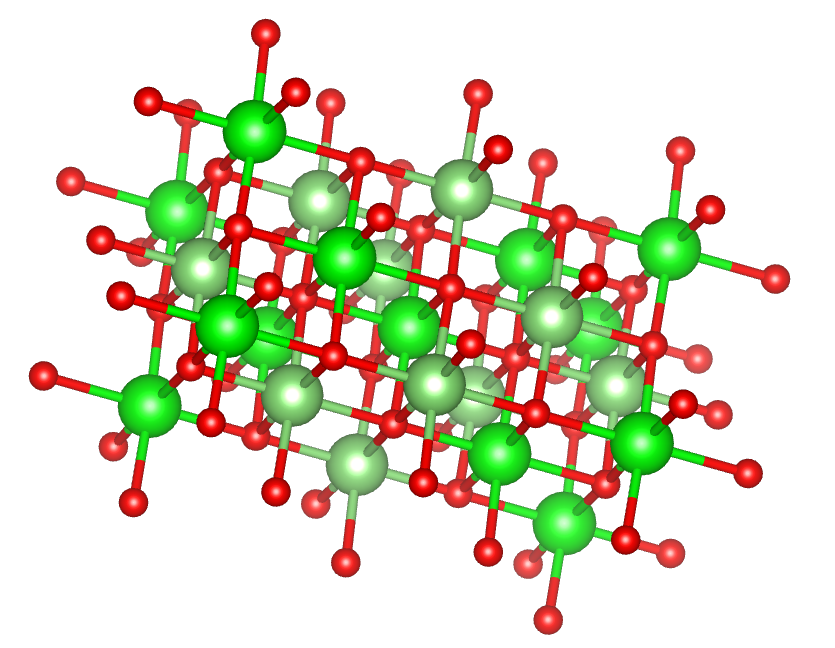}
      \caption{Ground truth}
      \label{fig:LiZrO2-a}
    \end{subfigure}\hfill
    \begin{subfigure}[b]{0.45\textwidth}
      \includegraphics[width=\textwidth]{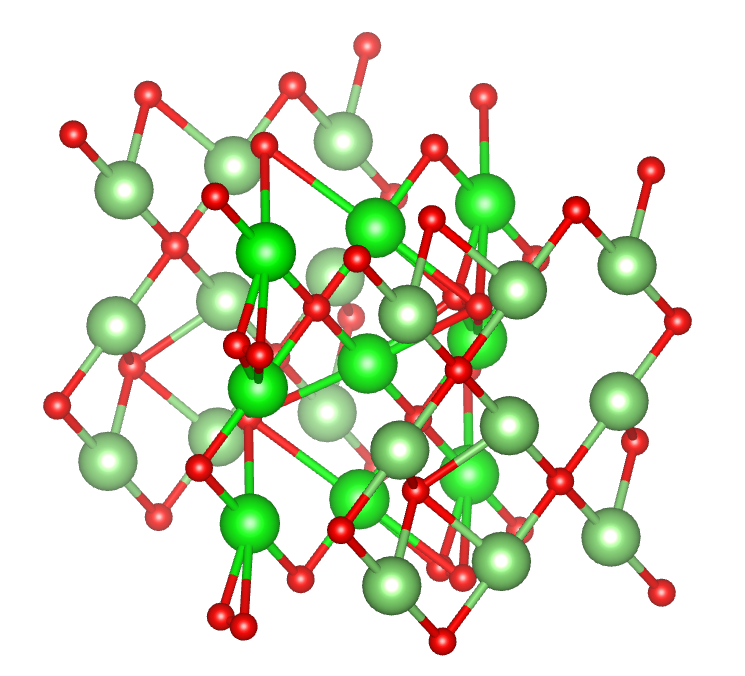}
      \caption{Predicted by RAS}
      \label{fig:LiZrO2-b}
    \end{subfigure}\\[0.5em]
    \begin{subfigure}[b]{0.45\textwidth}
      \includegraphics[width=\textwidth]{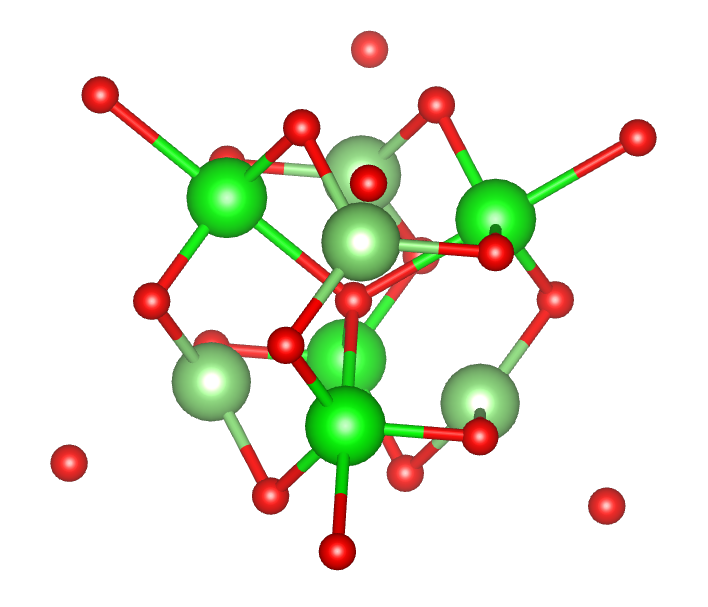}
      \caption{Predicted by BO}
      \label{fig:LiZrO2-c}
    \end{subfigure}\hfill
    \begin{subfigure}[b]{0.45\textwidth}
      \includegraphics[width=\textwidth]{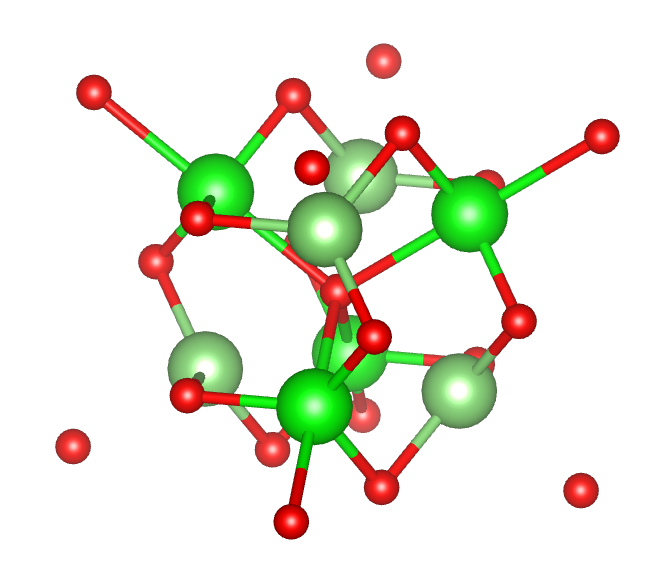}
      \caption{Predicted by PSO}
      \label{fig:LiZrO2-d}
    \end{subfigure}
    \captionof{figure}{Comparison of the ground truth and predicted crystal structures of LiZrO$_2$ by the RAS, BO, and PSO algorithms.}
    \label{fig:LiZrO2}
  \end{minipage}
\end{table}

\begin{table}[htbp]
  \centering
  \begin{minipage}{0.5\textwidth}
    \centering
    \caption{A metrics table generated by comparing the ground truth structure of LiTiSe$_2$O with the structures obtained using three different optimization algorithms from GN-OA: Random Acceleration Search (RAS), Particle Swarm Optimization (PSO), and Bayesian Optimization (BO).}
    \begin{tabular}{|llll|}
\hline
\multicolumn{4}{|c|}{\textbf{LiTiSe$_2$O}}                                                                                                                                 \\ \hline
\multicolumn{1}{|l|}{Algorithm}                                     & \multicolumn{1}{l|}{RAS}              & \multicolumn{1}{l|}{BO}               & PSO               \\ \hline
\multicolumn{1}{|l|}{Formation Energy}                              & \multicolumn{1}{l|}{-0.7530} & \multicolumn{1}{l|}{\textbf{-1.3047}} & -1.1161  \\ \hline
\multicolumn{1}{|l|}{\cellcolor[HTML]{FFFFFF}Energy Distance}       & \multicolumn{1}{l|}{93.8576} & \multicolumn{1}{l|}{\textbf{19.2273}} & 106.6080 \\ \hline
\multicolumn{1}{|l|}{\cellcolor[HTML]{FFFFFF}Wyckoff RMSE}          & \multicolumn{1}{l|}{0.3658}           & \multicolumn{1}{l|}{\textbf{0.3464}}           & 0.3958            \\ \hline
\multicolumn{1}{|l|}{\cellcolor[HTML]{FFFFFF}Wyckoff MAE}           & \multicolumn{1}{l|}{0.2669}           & \multicolumn{1}{l|}{\textbf{0.2458}}           & 0.3240            \\ \hline
\multicolumn{1}{|l|}{\cellcolor[HTML]{FFFFFF}Anonymous RMS}         & \multicolumn{1}{l|}{N/A}             & \multicolumn{1}{l|}{N/A}             & N/A              \\ \hline
\multicolumn{1}{|l|}{\cellcolor[HTML]{FFFFFF}RMS Distance}          & \multicolumn{1}{l|}{N/A}             & \multicolumn{1}{l|}{N/A}             & N/A     \\ \hline
\multicolumn{1}{|l|}{\cellcolor[HTML]{FFFFFF}Sinkhorn Distance}     & \multicolumn{1}{l|}{31.9493} & \multicolumn{1}{l|}{\textbf{23.6450}} & 30.4329  \\ \hline
\multicolumn{1}{|l|}{\cellcolor[HTML]{FFFFFF}Chamfer Distance}      & \multicolumn{1}{l|}{3.3169}  & \multicolumn{1}{l|}{\textbf{2.8741}}  & 3.5282   \\ \hline
\multicolumn{1}{|l|}{\cellcolor[HTML]{FFFFFF}Hausdorff Distance}    & \multicolumn{1}{l|}{3.2332}  & \multicolumn{1}{l|}{\textbf{2.6054}}  & 2.8125   \\ \hline
\multicolumn{1}{|l|}{\cellcolor[HTML]{FFFFFF}Superpose RMSD}        & \multicolumn{1}{l|}{\textbf{6.7413}}  & \multicolumn{1}{l|}{7.3776}  & 7.1555   \\ \hline
\multicolumn{1}{|l|}{Edit Graph Distance}                           & \multicolumn{1}{l|}{\textbf{12}} & \multicolumn{1}{l|}{15}          & 14  \\ \hline
\multicolumn{1}{|l|}{\cellcolor[HTML]{FFFFFF}FingerPrint Distance}  & \multicolumn{1}{l|}{2.6597}           & \multicolumn{1}{l|}{\textbf{2.0178}}  & 2.1986   \\ \hline
\multicolumn{1}{|l|}{\cellcolor[HTML]{FFFFFF}XRD Spectrum Distance} & \multicolumn{1}{l|}{1.9977}  & \multicolumn{1}{l|}{\textbf{1.2185}}  & 1.7125   \\ \hline
\multicolumn{1}{|l|}{\cellcolor[HTML]{FFFFFF}OFM Distance}          & \multicolumn{1}{l|}{0.4486}           & \multicolumn{1}{l|}{0.6285}  & \textbf{0.4263}   \\ \hline
\end{tabular}

\label{tab:LiTiSe2O}
  \end{minipage}%
  \hfill
  \begin{minipage}{0.45\textwidth}
    \centering
    \begin{subfigure}[b]{0.45\textwidth}
      \includegraphics[width=\textwidth]{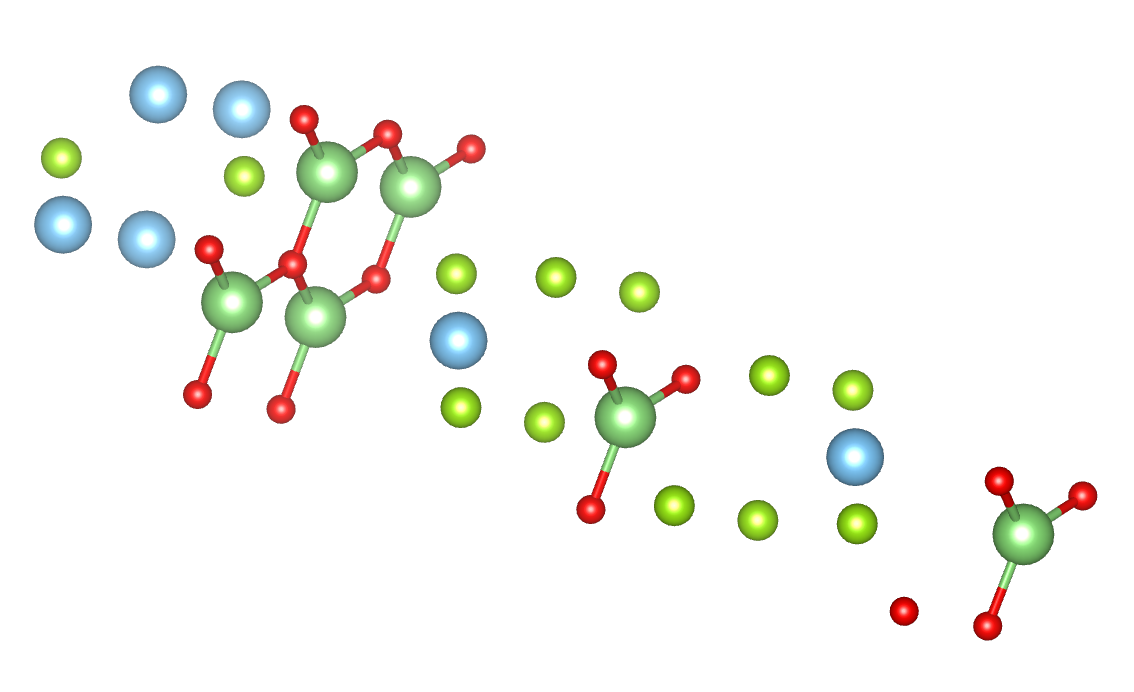}
      \caption{Ground truth}
      \label{fig:LiTiSe2O-a}
    \end{subfigure}\hfill
    \begin{subfigure}[b]{0.45\textwidth}
      \includegraphics[width=\textwidth]{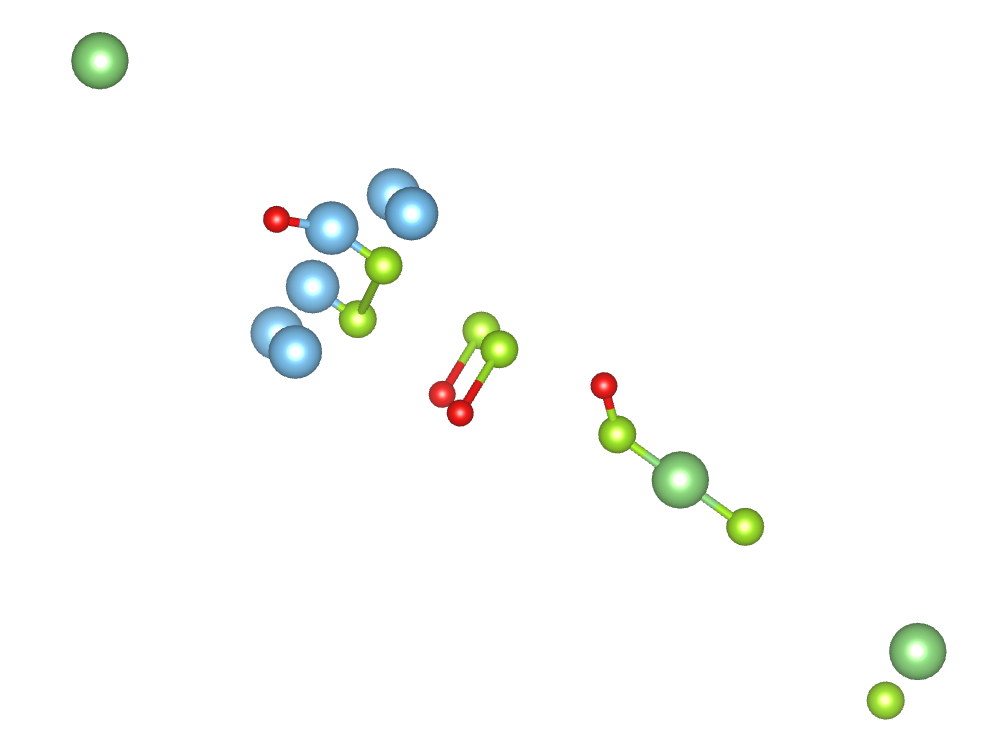}
      \caption{Predicted by RAS}
      \label{fig:LiTiSe2O-b}
    \end{subfigure}\\[0.5em]
    \begin{subfigure}[b]{0.45\textwidth}
      \includegraphics[width=\textwidth]{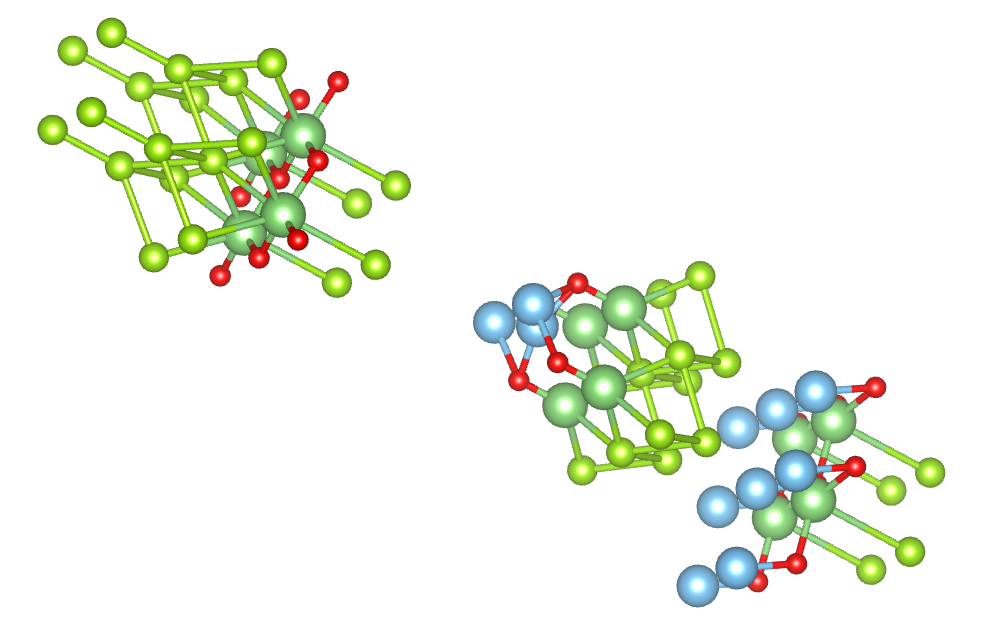}
      \caption{Predicted by BO}
      \label{fig:LiTiSe2O-c}
    \end{subfigure}\hfill
    \begin{subfigure}[b]{0.45\textwidth}
      \includegraphics[width=\textwidth]{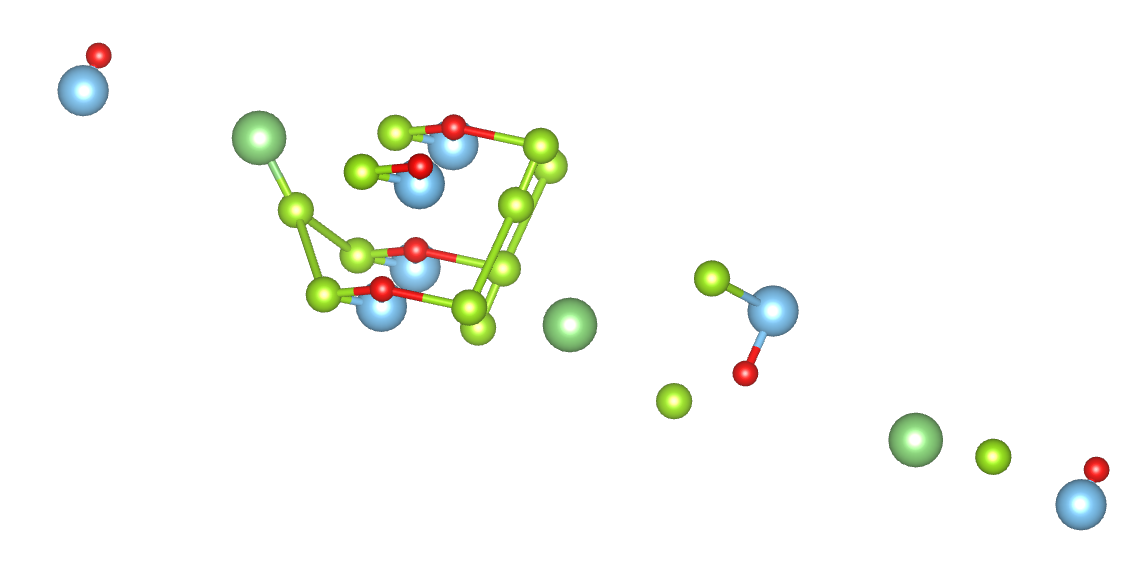}
      \caption{Predicted by PSO}
      \label{fig:LiTiSe2O-d}
    \end{subfigure}
    \captionof{figure}{Comparison of the ground truth and predicted crystal structures of LiTiSe$_2$O by the RAS, BO, and PSO algorithms.}
    \label{fig:LiTiSe2O}
  \end{minipage}
\end{table}

\FloatBarrier

\subsection{Trajectory studies of the GN-OA search algorithms in CSP of targets without polymorph structures}

\begin{figure}[ht!] 
 \begin{subfigure}[t]{0.33\textwidth}
        \includegraphics[width=\textwidth]{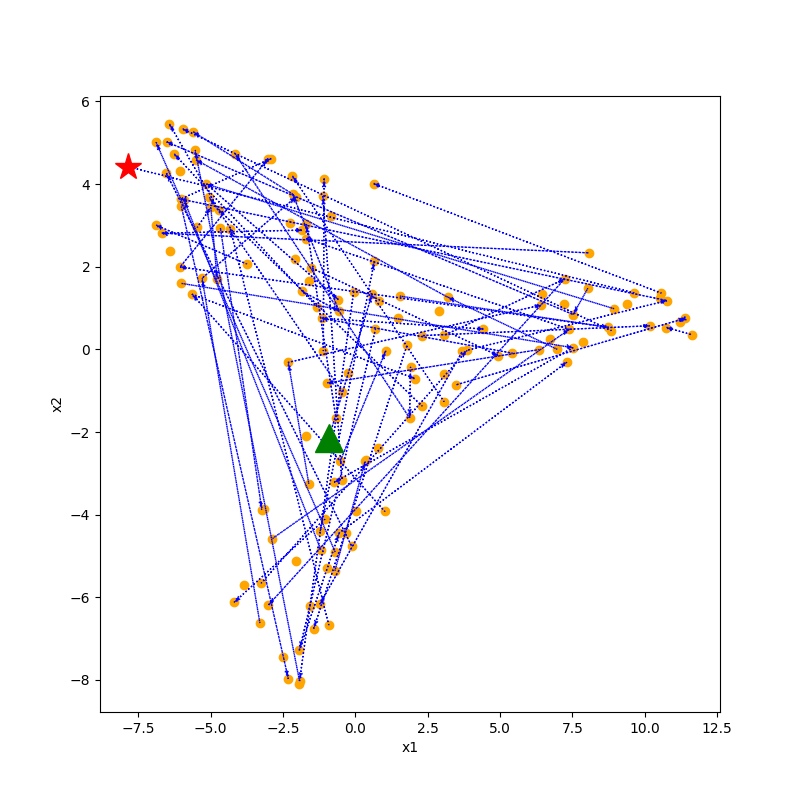}
        \caption{Random search}
        \vspace{-3pt}
        \label{fig:SrTiO3_target}
    \end{subfigure}
    \begin{subfigure}[t]{0.33\textwidth}
        \includegraphics[width=\textwidth]{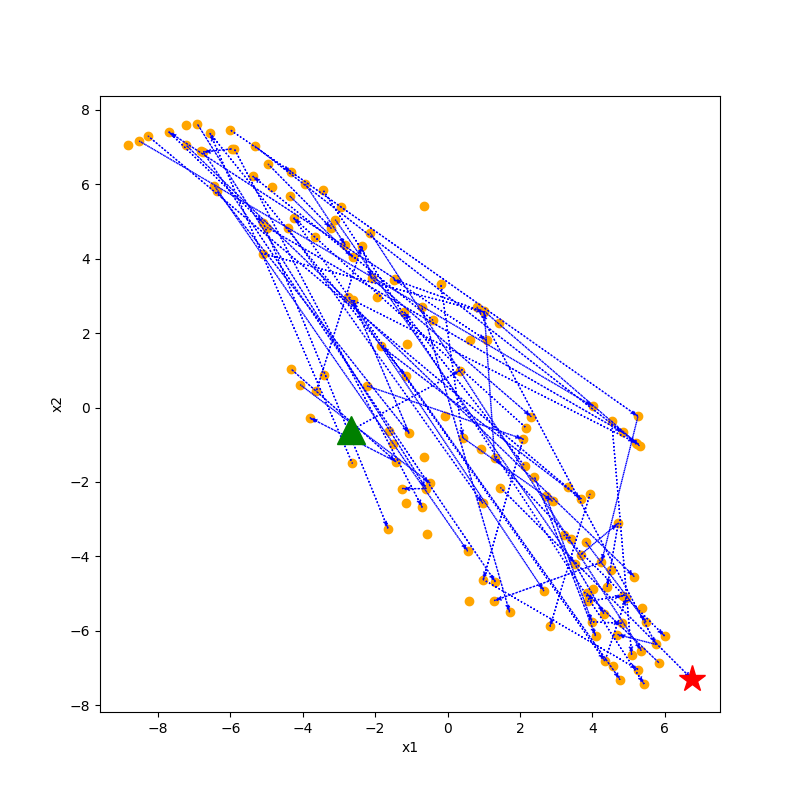}
        \caption{Bayesian optimization}
        \vspace{-3pt}
        \label{fig:Ni3S4_target}
    \end{subfigure}    
    \begin{subfigure}[t]{0.33\textwidth}
        \includegraphics[width=\textwidth]{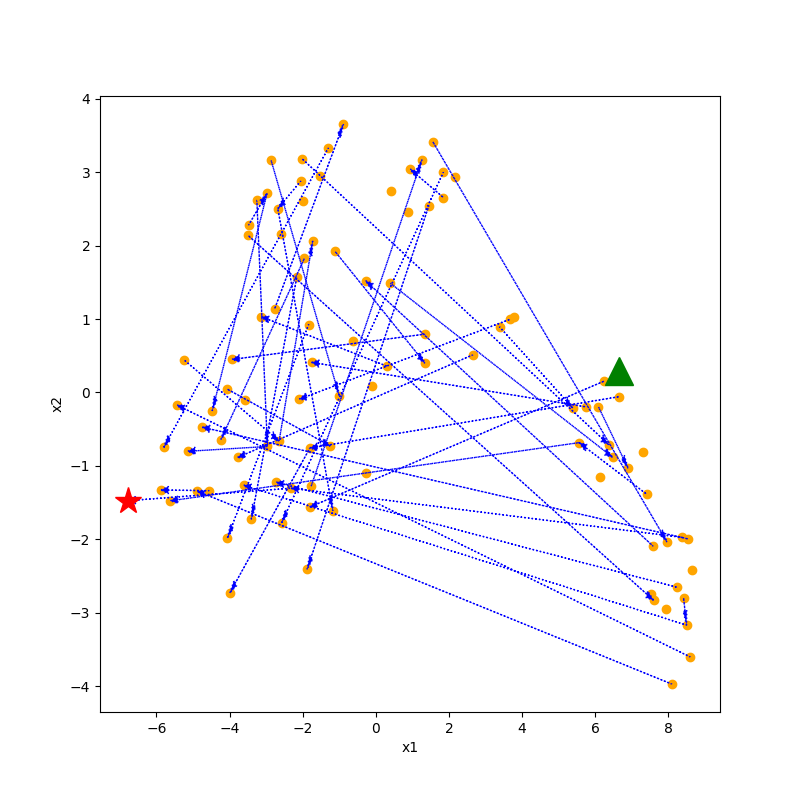}
        \caption{PSO}
        \vspace{-3pt}
        \label{fig:NiS2_target}
    \end{subfigure}    
   \caption{Trajectories of three search algorithms in crystal structure prediction of Ca$_4$S$_4$: (a) Random search with 5000 structure generation steps. 147 valid structures found; (b) Bayesian Optimization with 1000 structure generations. 140 valid structures found; (c) PSO with 5000 structure generations with 100 valid structures. The green triangles indicate the starting points while the red star indicates the ground truths.}
  \label{fig:traj1}
\end{figure}

Here we exploit the multi-dimensional CSP performance metrics to investigate the search behavior of three optimization algorithms used in the GN-OA CSP package \cite{cheng2022crystal}. We applied their random search, Bayesian optimization (BO), and particle swarm optimization (PSO) to search the structures of Ca$_4$S$_4$ and Ba$_3$Na$_3$Bi$_3$. For Ca$_4$S$_4$, we allocated 5,000 structure generations for the random search algorithm which generated 147 valid structures (readable by Pymatgen). We allocated 1,000 structure generation steps for the BO algorithm which traversed 140 valid structures. For PSO, we used 5,000 structure generation steps which created only 100 valid structures over the search process. For all the valid structures during a search, we calculated their distance metrics to the ground truth structure and then mapped the distance features using t-SNE to 2D dimension points. We then plot the trajectory of the structure search over time by connecting two consecutive points if the newer structure has lower energy than the previous one. The results are shown in Figure \ref{fig:traj1}. Note the green triangles indicate the starting points while the red stars represent the ground truth structures. 

First, we found that for all three algorithms, it is challenging for them to generate valid structures during their search. Both Random search and PSO only generated less than 150 structures for a total of 5,000 structure generations. In comparison, the BO algorithm generated 140 valid structures with only 1,000 structure generations. This is consistent with the authors' observation of GN-OA that the BO has better performance in their CSP experiments. From Figure \ref{fig:traj1} (a) and (c), we found that the Random search and PSO algorithms tend to jump around in a larger design space. In contrast, the BO algorithm is more focused during its search (Figure \ref{fig:traj1}(b)).

\begin{figure}[ht!] 
 \begin{subfigure}[t]{0.33\textwidth}
        \includegraphics[width=\textwidth]{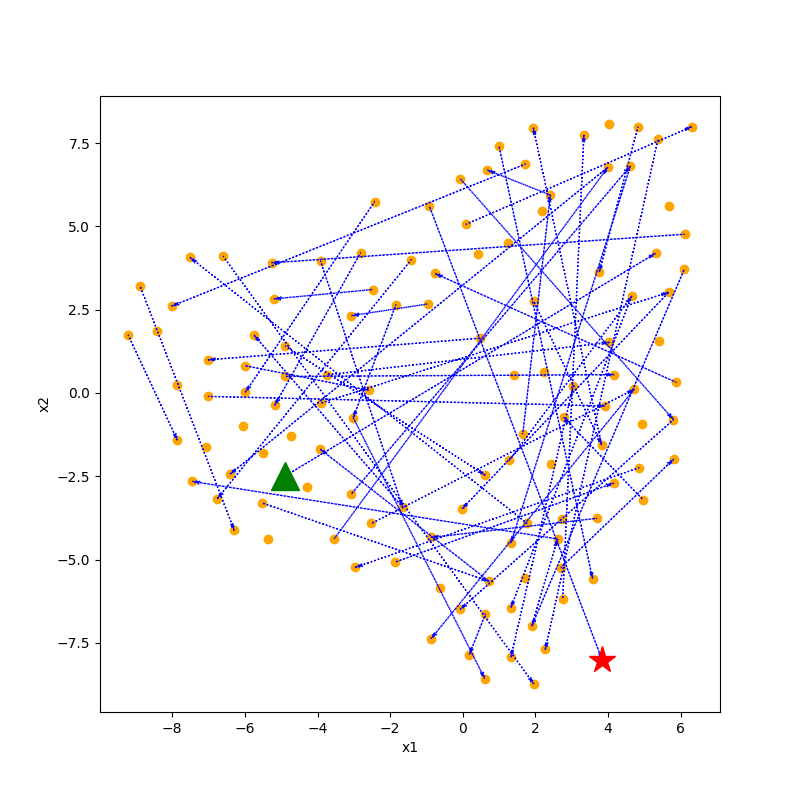}
        \caption{Random search}
        \vspace{-3pt}
        \label{fig:SrTiO3_target}
    \end{subfigure}
    \begin{subfigure}[t]{0.33\textwidth}
        \includegraphics[width=\textwidth]{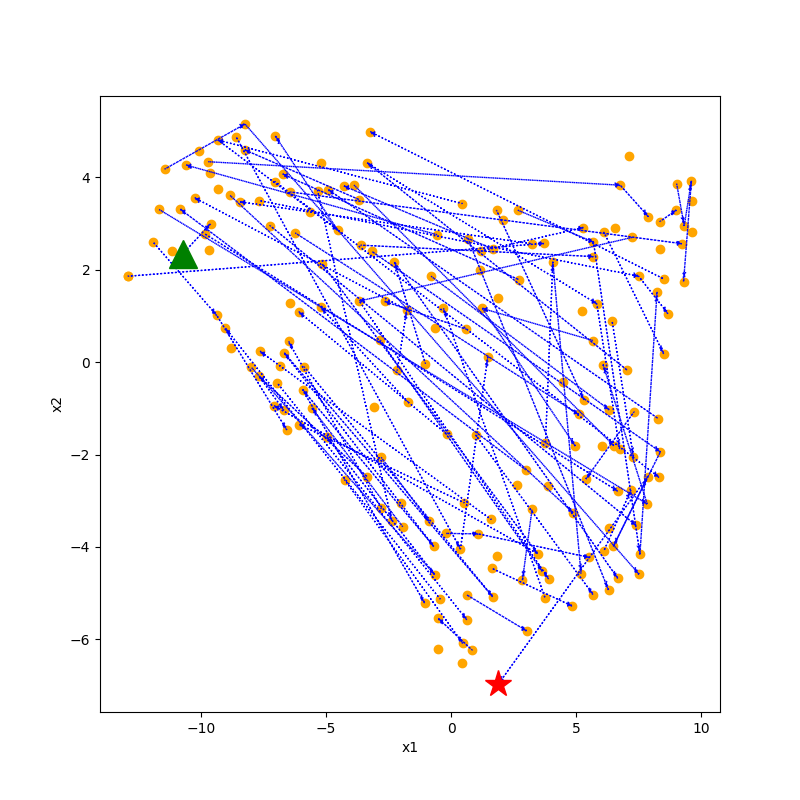}
        \caption{Bayesian optimization}
        \vspace{-3pt}
        \label{fig:Ni3S4_target}
    \end{subfigure}    
    \begin{subfigure}[t]{0.33\textwidth}
        \includegraphics[width=\textwidth]{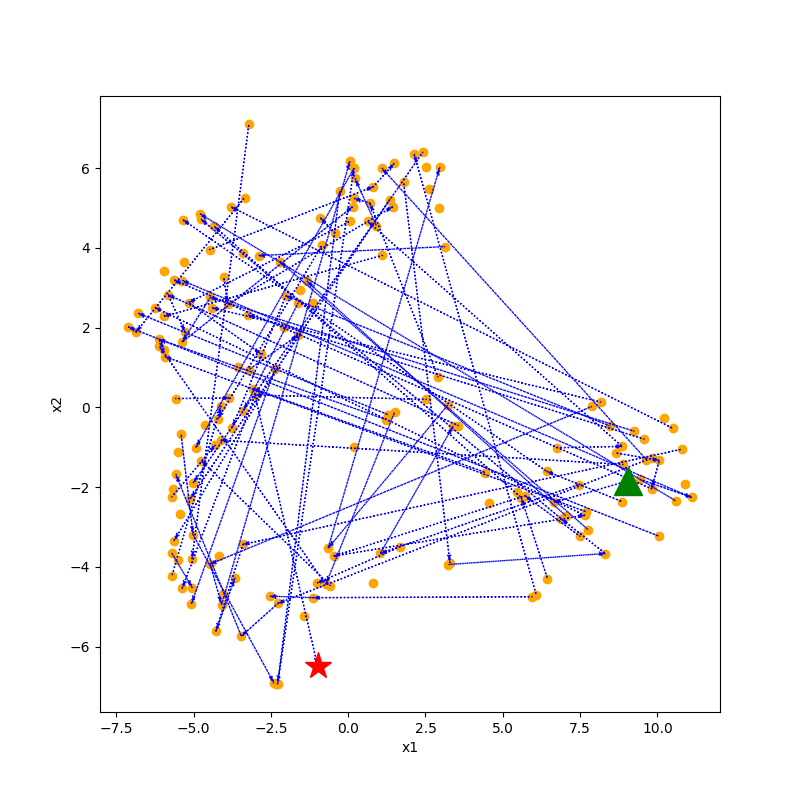}
        \caption{PSO}
        \vspace{-3pt}
        \label{fig:NiS2_target}
    \end{subfigure}    
   \caption{Trajectories of three search algorithms in crystal structure prediction of Ba$_3$Na$_3$Bi$_3$: (a) Random search with 50000 structure generation steps. 121 valid structures found; (b) Bayesian Optimization with 3000 structure generations. 195 valid structures found; (c) PSO with 6000 structure generations with 177 valid structures. }
  \label{fig:traj2}
\end{figure}

We further applied the three search algorithms to the structure prediction of a ternary compound Ba$_3$Na$_3$Bi$_3$, which is more challenging than Ca$_4$S$_4$. Figure \ref{fig:traj2} shows the three trajectories of the algorithms. First, we found that all three algorithms are much more difficult to generate valid structures, especially for the Random algorithm, which generates only 121 valid structures during its 50,000 tries. In contrast, the BO and PSO both generate 195 and 177 valid structures during their 3,000 and 6,000 structure generation steps, though the success rates of structure generation are still very low. The search trajectory patterns of the Random search and PSO are still more similar by jumping around a large area while the BO algorithm is more focused on their structure search. But compared to Figure \ref{fig:traj1} (b), the structure range  in Figure \ref{fig:traj2} is larger due to the higher complexity of the target structure Ba$_3$Na$_3$Bi$_3$. Our CSP metrics based trajectory analysis shows that current algorithms need to significantly improve their structure generation success rate to achieve higher efficiency in CSP.

\FloatBarrier

\FloatBarrier

\section{Conclusion}

Due to the complexity of structural changes during the search process of crystal structure algorithms, it is very difficult to measure the similarity of the candidate structures to the ground truth, especially when the algorithms cannot find the exact solution. This is especially challenging when the candidate structure and the target structure can have different spatial symmetries (space groups). We find that it is infeasible to use a single structure similarity measure to describe the CSP prediction quality of different algorithms. By evaluating a set of 9 structure distance measures (which we call as CSPMetrics), we find that using them together allows us to achieve a quantitative method to measure the prediction quality of predicted crystal structures compared to the ground truths. Application of our CSPmetric set has allowed us to gain interesting analysis of the structures during the search process of different CSP algorithms. While there are definitely rooms to further improve the metrics so that they can capture the prediction errors happening during CSP algorithm search, we believe our current CSPMetrics can be used as a good starting point to characterize benchmark different CSP algorithms. The availability of the source code additionally makes it easy for such evaluations.

\section{Data and Code Availability}

The test crystal structures are downloaded from the Materials Project database at \url{http://www.materialsproject.org}. The source code can be found at \url{https://github.com/usccolumbia/CSPBenchMetrics}

\section{Contribution}
Conceptualization, J.H.; methodology,J.H. L.W.,Q.L.,S.O.; software, L.W.,J.H., Q.L.; resources, J.H.; writing--original draft preparation, J.H., L.W., Q.L., S.O.; writing--review and editing, J.H., L.W.; visualization, J.H., L.W.; supervision, J.H.;  funding acquisition, J.H.

\section*{Acknowledgement}
The research reported in this work was supported in part by National Science Foundation under the grant 2110033. The views, perspectives, and content do not necessarily represent the official views of the NSF.

\bibliographystyle{unsrt}  
\bibliography{references}

\end{document}